\documentclass[preprint2]{aastex}

\newcommand{\bootes}{Bo\"otes}

\slugcomment{Preprint submitted to the ApJ}

\shorttitle{The Evolving Luminosity Function of Red Galaxies}
\shortauthors{Brown {\it et al.}}

\begin{document}

\title{The Evolving Luminosity Function of Red Galaxies}

\author{
Michael J. I. Brown\altaffilmark{1\dagger}, 
Arjun Dey\altaffilmark{2}, 
Buell T. Jannuzi\altaffilmark{2},
Kate Brand\altaffilmark{2}, 
Andrew J. Benson\altaffilmark{3,4}, 
Mark Brodwin\altaffilmark{5}
Darren J. Croton\altaffilmark{6}, 
Peter R. Eisenhardt\altaffilmark{5}
}
\altaffiltext{1}{Princeton University Observatory, Peyton Hall, Princeton, NJ 08544-1001, USA}
\altaffiltext{2}{National Optical Astronomy Observatory, Tucson, AZ 85726-6732, USA}
\altaffiltext{3}{Department of Physics, University of Oxford, Keble Road, Oxford OX1 3RH, UK}
\altaffiltext{4}{California Institute of Technology, 1200 E. California Blvd., Pasadena, CA 91125, USA}
\altaffiltext{5}{Jet Propulsion Laboratory, California Institute of Technology, 4800 Oak Grove Drive, Pasadena, CA 91109, USA}
\altaffiltext{6}{Department of Astronomy, University of California at Berkeley, Mail Code 3411, Berkeley, CA 94720, USA}
\altaffiltext{$\dagger$}{H.N. Russell Fellow}

\email{mbrown@astro.princeton.edu}

\begin{abstract}
We trace the assembly history of red galaxies since $z=1$, by measuring their 
evolving space density with the $B$-band luminosity function.
Our sample of 39599 red galaxies, selected from $6.96~{\rm deg}^2$ 
of imaging from the NOAO Deep Wide-Field and {\it Spitzer} IRAC Shallow surveys, 
is an order of magnitude larger, in size and volume, than comparable samples in the literature.
We measure a higher space density of $z\sim 0.9$ red galaxies than some of the recent 
literature, in part because we account for the faint yet significant galaxy flux 
which falls outside of our photometric aperture.
The $B$-band luminosity density of red galaxies, which effectively measures
the evolution of $\sim L^*$ galaxies, increases by only $36\pm 13\%$ from $z=0$ to $z=1$.
If red galaxy stellar populations have faded by  $\simeq 1.24$ $B$-band magnitudes since $z=1$,
the stellar mass contained within the red galaxy population has roughly doubled over the 
past $8~{\rm Gyr}$.  This is consistent with star-forming galaxies being transformed into 
$\lesssim L^*$ red galaxies after a decline in their star formation rates. In contrast, the 
evolution of $\simeq 4L^*$ red galaxies differs only slightly from a model with negligible $z<1$ 
star formation and no galaxy mergers. If this model approximates the luminosity evolution of red 
galaxy stellar populations, then $\simeq 80\%$ of the stellar mass contained within today's $4L^*$ 
red galaxies was already in place at $z=0.7$. While red galaxy mergers have been observed, such 
mergers do not produce rapid growth of $4L^*$ red galaxy stellar masses between $z=1$ and the present day.
\end{abstract}


\section{INTRODUCTION}
\label{sec:intro}

Red galaxies contain the majority of the stellar mass at low redshift \markcite{hog02}({Hogg} {et~al.} 2002), 
so understanding their formation and assembly is one of the key goals 
of both observational and theoretical extragalactic astronomy. 
The stellar populations of these galaxies are dominated by an
old component, with little ongoing star formation \markcite{tin68}(e.g., {Tinsley} 1968).
Observationally, it is unclear if this star formation occurred in situ at $z>1$, 
or if these stars formed in lower mass galaxies which were later assembled 
via galaxy mergers. Simulations of red galaxy evolution predict varying assembly
and star formation histories, with some papers concluding that assembly takes 
place primarily at high redshift \markcite{mez03,naa05}(e.g., {Meza} {et~al.} 2003; {Naab} {et~al.} 2005) while others suggest it continues today 
\markcite{del06}(e.g., {De Lucia} {et~al.} 2006).

The color-magnitude diagram of low redshift galaxies provides several important clues 
about the assembly of red galaxies. At low redshift there is a tight locus of red
galaxies in color-magnitude space \markcite{bow92,hog04,mci05}(e.g., {Bower}, {Lucey}, \& {Ellis} 1992; {Hogg} {et~al.} 2004; {McIntosh} {et~al.} 2005), with the most luminous red galaxies 
being slightly redder than less luminous red galaxies. This indicates that red galaxies 
of a given luminosity and metallicity have similar star formation histories, at least over 
the last few ${\rm Gyr}$. A red galaxy locus is also observed within the 
$z\sim 1$ cluster  and field galaxy populations \markcite{van00,bla03,bel04,wil05}({van Dokkum} {et~al.} 2000; {Blakeslee} {et~al.} 2003; {Bell} {et~al.} 2004; {Willmer} {et~al.} 2005).
The absence of very massive star-forming galaxies at low redshifts indicates that the most 
massive red galaxies are either formed at high redshift or they are assembled via mergers at $z<1$.
As the colors of red galaxies are a function of luminosity, mergers of red galaxies should
produce galaxies slightly blueward of the color-magnitude relation, and thus increase the scatter
within the relation.

Mergers of red galaxies, perhaps without significant merger-triggered star formation (dry mergers), 
have been observed at low redshift \markcite{lau88,van05}(e.g., {Lauer} 1988; {van Dokkum} 2005), but it is unclear how large
a role they play in galaxy formation. While there have been valiant attempts to measure the 
galaxy merger rate with redshift \markcite{lef00,bel06}(e.g., {Le F{\`e}vre} {et~al.} 2000; {Bell} {et~al.} 2006), there is debate about the 
selection function of merger candidates and the duration of observable galaxy mergers.
At low redshift, measured rates of red galaxy stellar mass growth via mergers span from 
$\simeq 1\%$ per ${\rm Gyr}$ \markcite{mas06}({Masjedi} {et~al.} 2006) to $\simeq 10\%$ per ${\rm Gyr}$ \markcite{van05}({van Dokkum} 2005).
 
The space density of red galaxies with redshift is a relatively direct measure of
the galaxy assembly history, and has fewer uncertainties than measurements of the 
galaxy merger rate. At the redshifts where galaxy assembly takes place, the space
density of galaxies as a function of stellar mass will evolve with redshift.
While conceptually simple, it is difficult to measure the galaxy space density 
accurately, as $z=1$ $L^*$ red galaxies are optically faint and their 
strong spatial clustering \markcite{bro03}(e.g., {Brown} {et~al.} 2003) results in significant cosmic variance.
Despite these difficulties, several groups have used measurements of the 
space density of galaxies with redshift to constrain red galaxy evolution and assembly
\markcite{lil95,lin99,che03,bel04,bun05,wil05,wak06}(e.g., {Lilly} {et~al.} 1995; {Lin} {et~al.} 1999; {Chen} {et~al.} 2003; {Bell} {et~al.} 2004; {Bundy} {et~al.} 2005; {Willmer} {et~al.} 2005; {Wake} {et~al.} 2006).
\markcite{fab05}{Faber} {et~al.} (2005) provides a useful summary of prior studies, and describes their varying conclusions.
The prior literature provides several plausible scenarios for $z<1$ red 
galaxy evolution, including passive evolution, assembly via dry mergers,
formation from fading blue galaxies, or a combination of the above.

This is the first in a series of papers which will discuss 
the assembly and evolution of $z<1$ red galaxies detected by the 
multiwavelength imaging surveys of the entire \bootes~ field. 
Other papers in this series will measure the stellar mass function, 
spatial clustering, and AGN content of red galaxies.  In this paper 
we present the evolving $B$-band luminosity function of $0.2<z<1.0$ 
red galaxies, and discuss the resulting implications for red galaxy 
evolution and assembly. While the rest-frame $B$-band is not ideal for 
tracing stellar mass, it has the advantages of being within the 
well studied optical wavelength range and remaining within the observed 
$B_WRI$ passbands of the NOAO Deep Wide-Field Survey (NDWFS) for 
redshifts of $z<1$.

The structure of this paper is as follows. We provide 
an overview of the NDWFS and {\it Spitzer} IRAC Shallow Survey in 
\S\ref{sec:surveys}. We discuss our catalogs, photometry, photometric redshifts, 
and galaxy rest-frame properties in \S\ref{sec:catalogs}. In \S\ref{sec:sample}
we describe the selection of our red galaxy sample. In \S\ref{sec:lf} we present
rest-frame $B$-band luminosity functions of red galaxies and compare our measurements with 
simple galaxy evolution models. In \S\ref{sec:litcomp} we compare our results with 
recent red galaxy surveys, and discuss discrepancies between various studies.
Our principal results and conclusions are summarized in \S\ref{sec:sum}.
Throughout this paper we use Vega based magnitudes and adopt a cosmology of 
$\Omega_m=0.3$, $\Omega_\Lambda=0.7$, $H_0=70~{\rm km}~{\rm s}^{-1}~{\rm Mpc}^{-1}$, 
$h=0.7$, and $\sigma_8=0.8$.

\section{THE SURVEYS}
\label{sec:surveys}

\subsection{The NOAO Deep Wide-Field Survey}

The NOAO Deep Wide-Field Survey (NDWFS) is an optical ($B_WRI$) and 
near-infrared ($K$) imaging survey of two $\approx 9.3~{\rm deg}^2$ 
fields with telescopes of the National Optical Astronomy Observatory \markcite{jan99}({Jannuzi} \& {Dey} 1999).
A thorough description of the observing strategy and data reduction
will be provided by B.~T.~Jannuzi et al. and A.~Dey et al. (both in preparation).
We utilize the third NDWFS data release\footnote{Available from the NOAO Science 
Archive at http://www.archive.noao.edu/ndwfs/} of optical imaging with the MOSAIC-I camera on the Kitt 
Peak 4-m telescope.
To obtain accurate optical colors with fixed aperture photometry across the \bootes~field, we have
smoothed copies of the released images so the stellar Point Spread Function (PSF) is a Moffat profile
with a full width at half maximum of $1.35^{\prime\prime}$ and $\beta=2.5$. 
The subfield  ${\rm NDWFSJ1428+3531}$ (each subfield is roughly one MOSAIC-I 
field-of-view) has poor seeing in the $B_W$ and $R$-bands, and has been excluded from this study.

\subsection{The IRAC Shallow Survey}

The IRAC instrument \markcite{faz04}({Fazio} {et~al.} 2004) on the {\it Spitzer} Space Telescope provides 
simultaneous broad-band images at $3.6$, $4.5$, $5.8$, and $8.0~\mu {\rm m}$.
The IRAC Shallow Survey, a guaranteed-time observation program of the 
IRAC instrument team, covers $8.5~{\rm deg^2}$ of \bootes~ with three or more 
30 second exposures per position. \markcite{eis04}{Eisenhardt} {et~al.} (2004) present an overview of the 
survey design, reduction, calibration, and preliminary results.
Despite the short integration times, the IRAC Shallow Survey  
easily detects $z\sim 1.4$ cluster galaxies \markcite{sta05,els06,brod06}({Stanford} {et~al.} 2005; {Elston} {et~al.} 2006; {Brodwin} {et~al.} 2006).
In this paper we utilize the $3.6$ and $4.5~\mu {\rm m}$ imaging 
to remove contaminants (e.g., stars, quasars) from our galaxy sample and for empirical 
photometric redshifts. The observed $I-[3.6]$ color of $z<1$ red galaxies 
is a superb redshift indicator, due to the redshifting of the $1.6~\mu{\rm m}$ 
${\rm H}^-$ stellar opacity feature \markcite{sim99,saw02}(e.g., {Simpson} \& {Eisenhardt} 1999; {Sawicki} 2002).

\section{THE OBJECT CATALOG}
\label{sec:catalogs}

\subsection{OBJECT DETECTION}

We detected sources using SExtractor $2.3.2$ \markcite{ber96}({Bertin} \& {Arnouts} 1996), run in single-image 
mode on the  $I$-band images of the NDWFS third data release. 
The NDWFS \bootes~field comprises of 27 optical subfields, each of which are $35^{\prime} \times 35^{\prime}$
in size, that are are designed to have overlaps of several arcminutes. Individual objects can 
be detected in multiple subfields, which is particularly useful for photometric calibration, 
though the effective exposure time generally decreases towards subfield edges.
For this paper, we only include objects which are detected within 
nominal subfield boundaries, which we define to have overlaps of only
tens of arcseconds. In these small overlap regions, we search for objects with 
$I$-band detections in multiple subfields and retain the detection
with the highest quality data (i.e., without bad pixels or 
with the longest exposure time). To minimize the number of faint spurious
galaxies in our catalog, we exclude regions surrounding very extended galaxies 
and saturated stars. We also exclude regions that do not have good coverage
from both the NDWFS and the IRAC Shallow Survey. The final sample area
is $6.96~{\rm deg}^2$ over a $2.7^\circ \times 3.3^\circ$ field-of-view.

\subsection{PHOTOMETRY}

We used our own code to measure aperture photometry for each object 
in the optical and IRAC passbands. 
To reduce contaminating flux from neighboring objects, we used 
SExtractor segmentation maps to exclude pixels associated with 
neighboring objects detected in any of the three optical bands. We corrected 
the aperture photometry for missing pixels (from the segmentation maps 
or bad pixels) by using the mean flux per pixel measured in a series on 
$0.5^{\prime\prime}$ wide annuli surrounding each object position. 
Accurate $1\sigma$ uncertainties for the photometry were determined by 
measuring photometry at $\simeq 100$ positions within $2^\prime$ of the 
object position and finding the uncertainty which encompassed 68.7\% of the 
measurements. We also use the median of the fluxes measured at these positions
to subtract small errors which could be present in our sky background estimate.
Typical random uncertainties for $I$-band $4^{\prime\prime}$ diameter aperture 
photometry are $0.03$ magnitudes at $I=21$, increasing to $0.2$ magnitudes at $I=23$.
Typical random uncertainties for $3.6~\mu{\rm m}$ $4^{\prime\prime}$ aperture 
photometry are $0.02$ magnitudes at $[3.6]=16$, increasing to $0.2$ magnitudes at $[3.6]=19$.

\begin{figure*}[hbt]
\vspace{1cm}
\epsscale{0.80}
\plotone{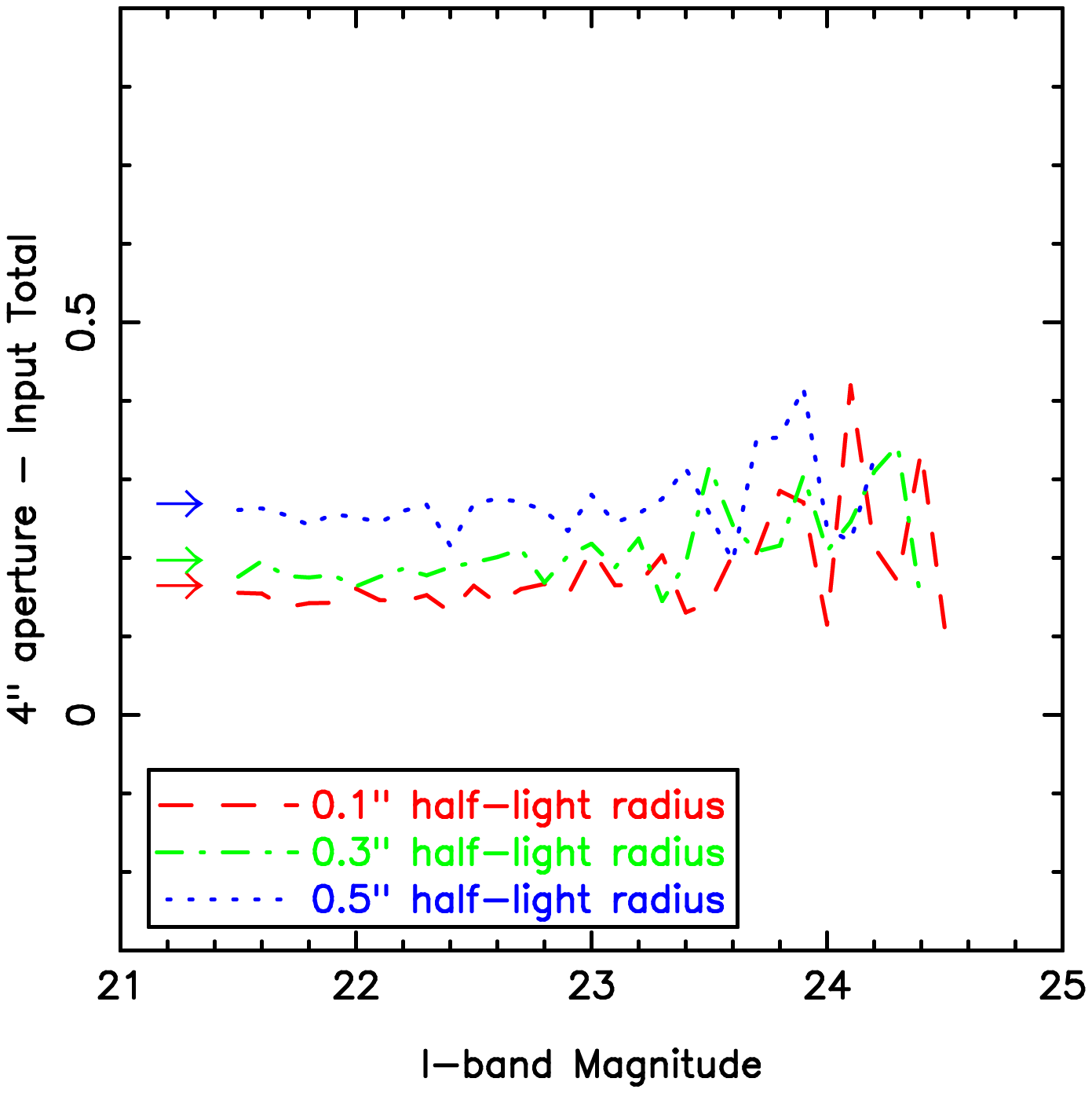}\plotone{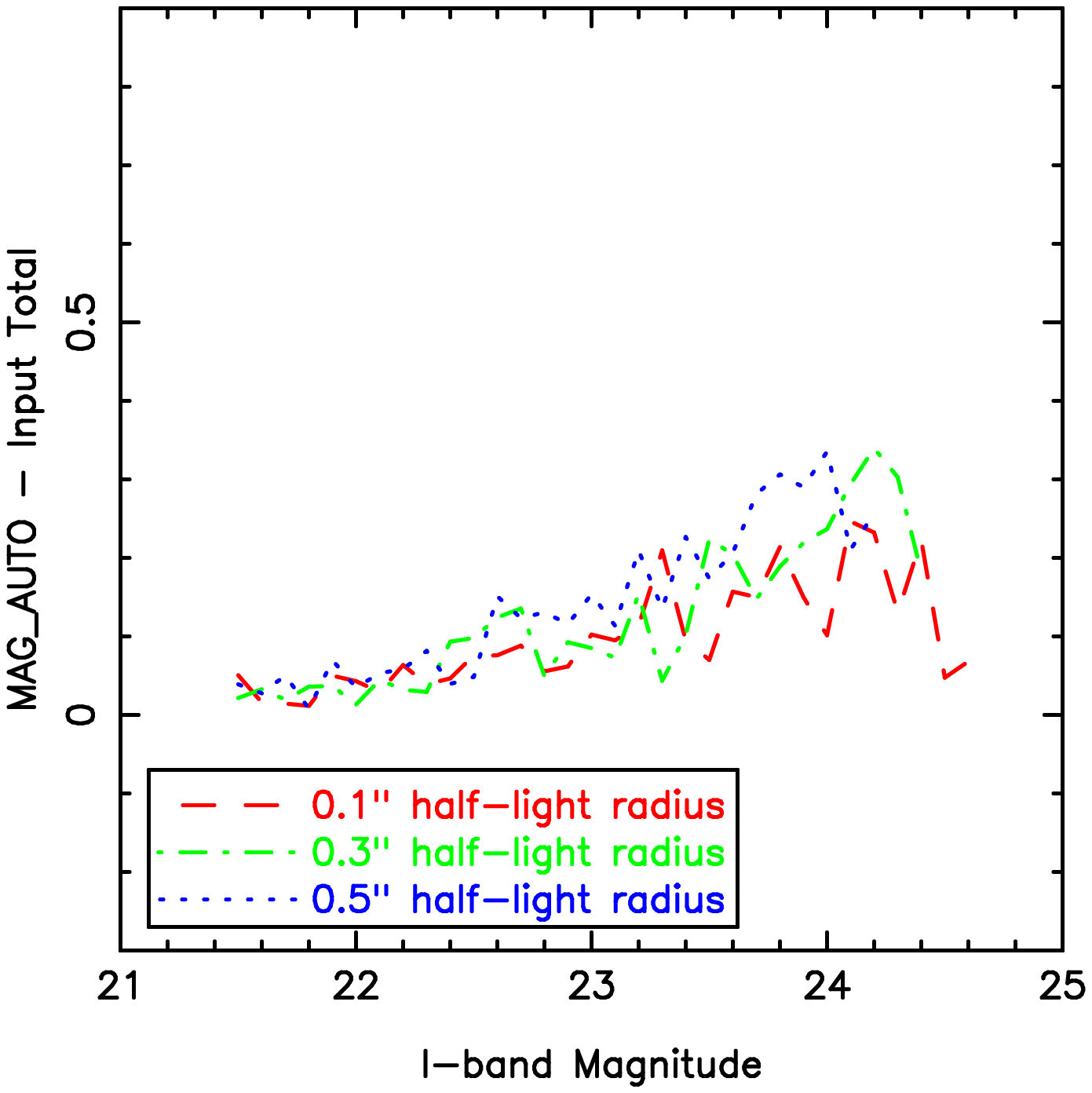}
\epsscale{1.00}
\caption{The offset between measured and total photometry for $4^{\prime\prime}$ 
aperture photometry (left) and SExtractor's MAG\_AUTO (right). 
We only plot galaxies over the apparent magnitude range where the completeness is 
50\% or more, and we treat non-detections as measurements of $I=99$.
The predicted offsets due to flux outside the $4^{\prime\prime}$ aperture are 
shown with arrows, and these are consistent with our measurements.
SExtractor's MAG\_AUTO photometry is closer to total magnitudes than 
$4^{\prime\prime}$ aperture photometry, but has systematics as a function of both 
half-light radius and apparent magnitude. Luminosity functions 
using MAG\_AUTO for total magnitudes may have systematics as a function of 
both luminosity and redshift.
\label{fig:sys}}
\end{figure*}

To verify the accuracy of our uncertainties and to search for systematic errors, 
we added artificial galaxies to our data, recovered them with SExtractor, 
and measured their photometry. The artificial galaxies have \markcite{dev48}{de Vaucouleurs} (1948) 
profiles truncated at seven half-light radii (though our results are not 
particularly sensitive to the truncation radius), which were then convolved with 
our PSF. Figure~\ref{fig:sys} plots the offset between total and measured
luminosities for these galaxies. This offset is due to flux outside outside of our
$4^{\prime\prime}$ diameter aperture, and is a predictable function of half-light radius.
The observed half-light radius is a function of both galaxy luminosity and redshift,
and we discuss corrections from $4^{\prime\prime}$ aperture photometry to total 
magnitudes in \S\ref{sec:rest}. After accounting for this predicted offset, 
we find that the difference between the input and measured photometry is within the quoted 
$1\sigma$ uncertainties $\simeq 70\%$ of the time.

It is useful to compare our photometry with SExtractor's ${\rm MAG\_AUTO}$, 
which is often used to measure ``total'' magnitudes in galaxy 
surveys \markcite{bel04,zuc05}(e.g., {Bell} {et~al.} 2004; {Zucca} {et~al.} 2005).
While SExtractor is an extremely useful tool for source detection, photometry and 
classification, it cannot be expected to provide perfect measurements of all objects in all surveys. 
Using the same artificial galaxies as described above, we find 80\% of the MAG\_AUTO 
measurements differ from the total magnitude by more than the quoted $1\sigma$ uncertainty. 
This is due to SExtractor's assumption that the sky background is Gaussian random noise without
source confusion, which is only a valid approximation for some imaging data.

We show in Figure~\ref{fig:sys} that MAG\_AUTO is closer to a total magnitude than 
$4^{\prime\prime}$ aperture photometry, though it has systematic errors which are
a function of apparent magnitude. In addition, its random errors are 40\% larger
than  $4^{\prime\prime}$ aperture photometry. The systematic error with magnitude  
may be due to the elliptical MAG\_AUTO aperture being defined using the second-order
image moments of object pixels above an isophotal threshold. It is therefore plausible
that the MAG\_AUTO aperture is slightly too small for relatively extended faint galaxies.
Luminosity functions using MAG\_AUTO for total magnitudes may have systematics of $\sim 0.2$
magnitudes at faint magnitudes, though the exact size of this offset will depend both on the
imaging data and user defined SExtractor parameters. Accurate luminosity functions require
corrections for galaxy flux which falls beyond the photometric aperture.

\subsection{PHOTOMETRIC REDSHIFTS}

We determined redshifts for our galaxies using the empirical 
ANNz photometric redshift code \markcite{fir03,col04}({Firth}, {Lahav}, \& {Somerville} 2003; {Collister} \& {Lahav} 2004). 
When a sufficiently large training set is available, empirical 
photometric redshifts are as precise or better than other 
techniques \markcite{csa03,brod06}(e.g., {Csabai} {et~al.} 2003; {Brodwin} {et~al.} 2006). 
This approach does restrict us to $z<1$, where thousands of spectroscopic 
redshifts are available in \bootes~ to calibrate the photometric redshifts.
ANNz uses artificial neural networks to determine the relationship between 
measured galaxy properties and redshift. It does not use any 
prior assumptions about the shape of galaxy spectral energy distributions (SEDs), 
though it does assume the relationship between galaxy properties and redshift is a
relatively smooth function. ANNz works best when the training set is a large
representative subset of the science galaxy sample. 

The basis of our training set is galaxies with spectroscopic redshifts
in the \bootes~field. The ongoing AGN and Galaxy Evolution 
Survey (AGES, C.~S.~Kochanek~et~al.~in preparation) of \bootes~has obtained 
spectroscopic redshifts of $\simeq 16000$ $I\lesssim 20$ galaxies, 
using the MMT's Hectospec multiobject spectrograph \markcite{fab98}({Fabricant} {et~al.} 2005).
At fainter magnitudes,  most of the spectroscopic redshifts are
from various spectroscopic campaigns with the W.~M.~Keck, Gemini, 
and Kitt Peak observatories, which have jointly resulted in redshifts 
for several hundred galaxies.

We trained and measured photometric redshifts using the $4^{\prime\prime}$
aperture photometry and the 2nd order moments of the $I$-band light distribution.
For training and determining photometric redshifts we used ${\rm asinh}$ magnitudes 
\markcite{lup99}({Lupton}, {Gunn}, \& {Szalay} 1999) rather than fluxes or conventional $2.5{\rm log}_{10}$ magnitudes. ANNz does not produce accurate 
photometric redshifts with fluxes that span a vast dynamic range, while conventional magnitudes
only measure positive fluxes. We only use  ${\rm asinh}$ magnitudes when estimating
ANNz photometric redshifts and use conventional magnitudes throughout the remainder of the paper.
The second order image moments are useful for reducing the number of bright galaxy photometric redshifts
with large errors. In particular, they reduce photometric redshift outliers produced by 
low redshift edge-on spirals whose dust lanes produce colors similar to higher redshift objects.
As we do not want the shape measurements to be a function of $I$-band depth, which varies
slightly across \bootes, we measure the 2nd order moments with pixels above an
$I$-band surface brightness of $22.5~{\rm mag}~{\rm arcsec}^{-2}$. 

We applied several restrictions to the training set to improve the 
photometric redshifts of $z<1$ red galaxies. We restricted the training 
set to $z<1.5$ spectroscopic galaxies, so $z>1.5$ galaxies did not perturb the ANNz
photometric redshift solution for $z<1.5$ galaxies. While we include red galaxies
with X-ray and radio counterparts in our final sample, these objects are overrepresented
in our spectroscopic samples and a small fraction of these sources could have unusual colors
due to the contribution of the AGN. We have therefore excluded objects with counterparts in the 
{\it Chandra} X\bootes~ survey \markcite{ken05,mur05,bra06}({Kenter} {et~al.} 2005; {Murray} {et~al.} 2005; {Brand} {et~al.} 2006) or the FIRST radio survey \markcite{bec95}({Becker}, {White}, \& {Helfand} 1995) from the
photometric redshift training set.
The small number of galaxies with unusual apparent colors which remained in the training 
set were removed with apparent color cuts which are a function of spectroscopic redshift.

\begin{figure*}[hbt]
\vspace{1cm}
\epsscale{0.70}
\plotone{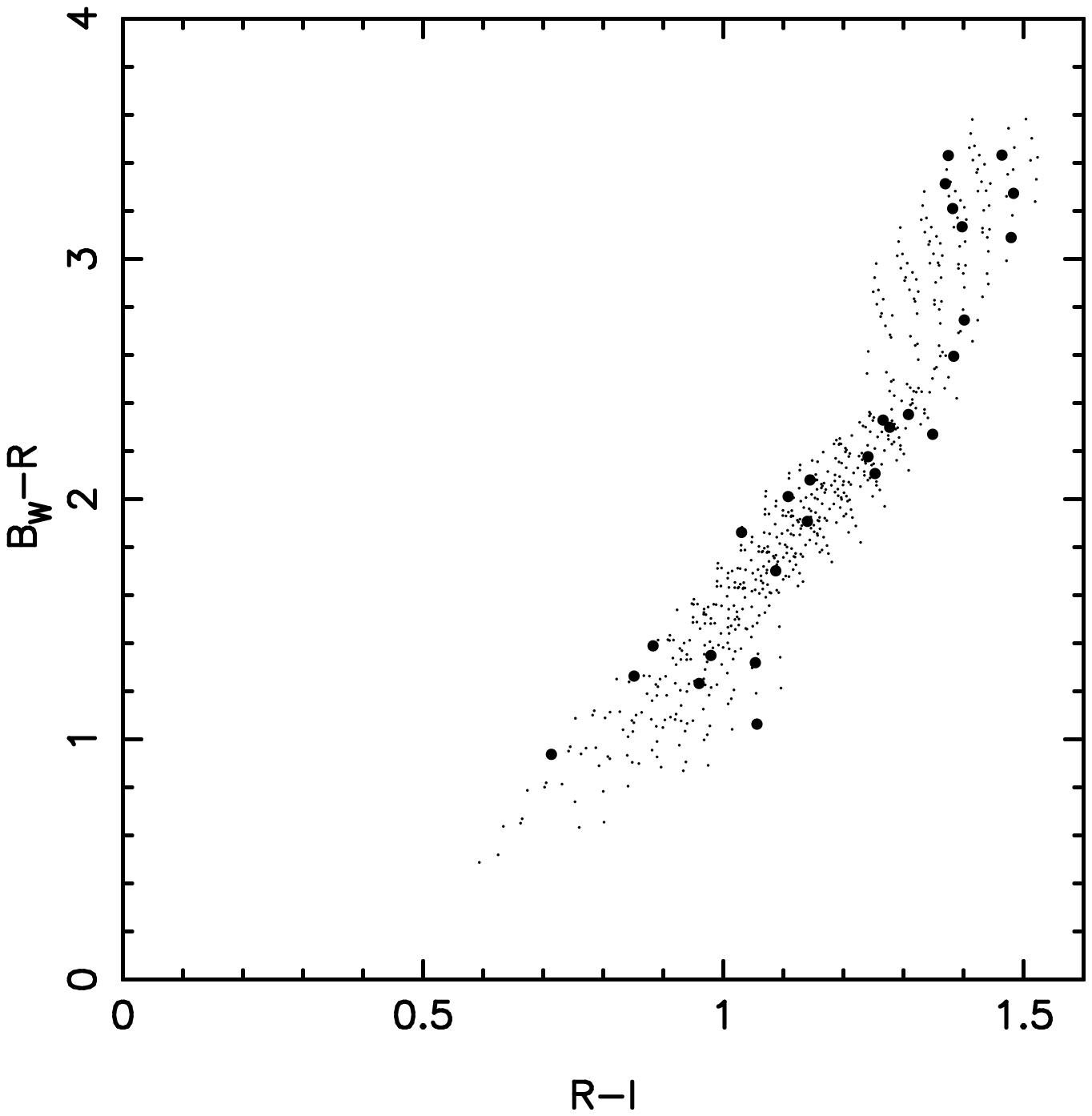}\plotone{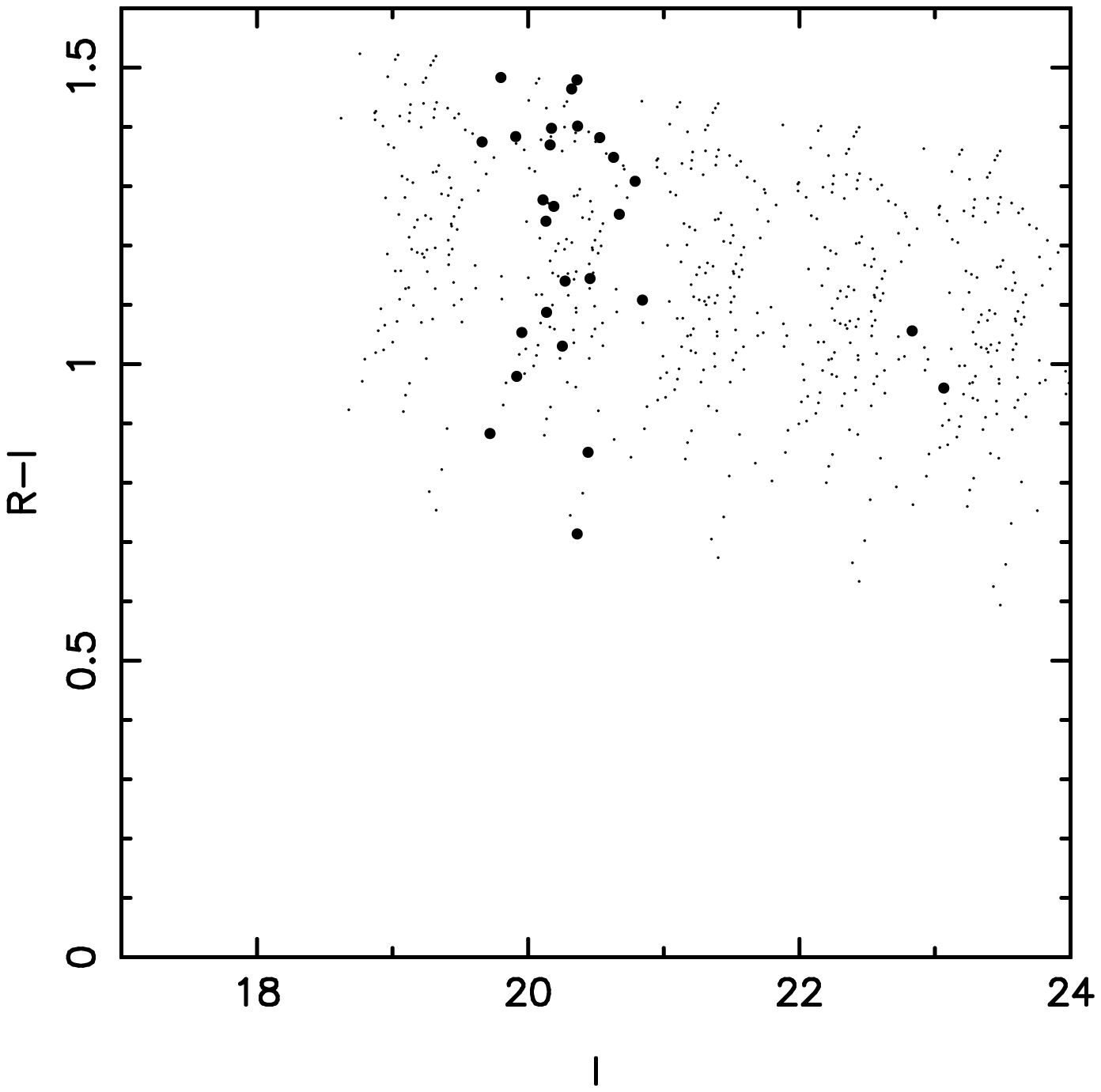}
\epsscale{1.00}
\caption{Apparent color-color and color-magnitude diagrams for photometric redshift training set 
galaxies with spectroscopic redshifts $0.78<z<0.82$.
Bold symbols denote real galaxies while artificial objects produced by interpolating in color or
extrapolating in magnitude are shown with dots. As the observed colors of galaxies are a strong 
function of redshift and a weak function of luminosity, approximations of the color-color and 
color-magnitude relations can be used to extrapolate the training set to faint magnitudes.
\label{fig:train}}
\end{figure*}

At $z>0.6$, there are only 280 red and blue galaxies fainter than $I=20.5$ with spectroscopic
redshifts in the training set, which is less than ideal for training ANNz. 
As the shape of the galaxy color-locus at any given redshift is relatively 
simple and smooth, we were able to increase the size of the training set by adding
interpolated objects to the training set. We did this when we found galaxies
within 0.02 in redshift and within 0.5 in $B_W-R$ color of each other.
As red galaxies at $z<1$ follow relatively tight 
color-magnitude \markcite{bow92,bel04,mci05}(e.g., {Bower} {et~al.} 1992; {Bell} {et~al.} 2004; {McIntosh} {et~al.} 2005) and size-luminosity
relations \markcite{ben92,she03}(e.g., {Bender}, {Burstein}, \& {Faber} 1992; {Shen} {et~al.} 2003), we were able to extrapolate the 
training set to fainter magnitudes by making faint copies of bright objects 
with slightly altered photometry and image size parameters. 
As the colors and sizes of galaxies are a very strong function of 
redshift, approximations of the color-magnitude and size-luminosity 
relations at a given redshift are sufficient for the photometric redshift 
training set. 
Figure~\ref{fig:train} plots the color-color and color-magnitude diagrams 
of training set galaxies at $z=0.80$, along with interpolated and extrapolated objects.

\begin{figure}
\plotone{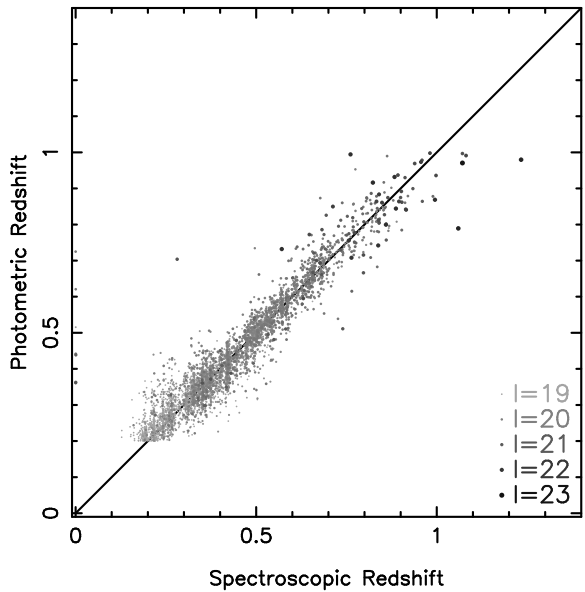}
\caption{Comparison of spectroscopic and photometric redshifts for the 
red galaxy sample. Symbol greyscale is a function of apparent magnitude, with 
the small number of $I>22$ galaxies with spectroscopic redshifts having large black symbols.
A total of 4314 objects are plotted, and only $0.3\%$ of these objects have errors
of $>0.2$ in redshift. 
\label{fig:photoz}}
\end{figure}

In Figure~\ref{fig:photoz} we plot the photometric and spectroscopic redshifts of
4314 objects which meet our red galaxy selection criteria (which we discuss in 
\S\ref{sec:sample}), including radio and X-ray sources. 
We also provide estimates of photometric redshift uncertainties for a series of 
redshift and apparent magnitude bins in Table~\ref{table:photoz}. 
The uncertainties were determined using the relevant percentiles, 
and do not assume a Gaussian distribution of photometric redshift errors. 
As the observed range of galaxy colors broadens with increasing apparent 
magnitude, at the very least due to photometric errors, the random errors of 
the photometric redshifts also increase with apparent magnitude. 
Measured systematic errors are much smaller than the random errors listed in 
Table~\ref{table:photoz}. 
The accuracy of our red galaxy photometric redshifts are comparable to the 
best broad-band photometric redshifts available in the literature \markcite{mob04}(e.g., {Mobasher} {et~al.} 2004). 
The $1\sigma$ uncertainties of our photometric redshifts are $\simeq 0.1$ in redshift 
at $I=22$, and this decreases to $\simeq 0.03$ at $I=19.5$. 

\subsection{REST-FRAME PROPERTIES}
\label{sec:rest}

To measure the rest-frame absolute magnitudes and colors of NDWFS galaxies, 
we used maximum likelihood fits of \markcite{bru03}{Bruzual} \& {Charlot} (2003) SED models to the 
$B_WRI$ $4^{\prime\prime}$ diameter aperture photometry.  Throughout this paper we use 
solar metallicity models with a \markcite{sal55}{Salpeter} (1955) initial mass function, a formation redshift 
of $z=4$, and exponentially declining star formation rates. 
These models provide a reasonable  approximation of the observed SEDs of red galaxies.
Unlike empirical templates, the \markcite{bru03}{Bruzual} \& {Charlot} (2003) models allow us to estimate stellar masses 
and star formation rates, which we will use in future papers (K.~Brand~et~al. and A.~Dey~et~al., 
both in preparation). For this paper, we will only use the rest-frame optical colors and absolute magnitudes. 

Other stellar population models can be fitted to red galaxies, though 
changing models has little impact on our $B$-band luminosity functions and our 
conclusions. Shifting the formation redshift results in different $\tau$ models 
being fitted to the red galaxies, but the model $B$-band luminosity evolution of 
red galaxy stellar populations at $z<1$ changes by  $0.1$ magnitudes or less.
Changing from a \markcite{sal55}{Salpeter} (1955) to a \markcite{cha03}{Chabrier} (2003) initial mass function decreases 
galaxy stellar masses by 25\%, but has little impact on the predicted evolution 
of galaxy optical colors and luminosities. Models with low metallicities are bluer
than the reddest galaxies, while models with 1.5 times Solar metallicity are offset 
from observed $B_WRI$ galaxy loci from $z=0.2$ to $z=1.0$. 

As with all current galaxy SED templates and models, there are small systematic 
errors. The $\tau$ models do not perfectly match the evolving rest-frame $U-V$ colors 
of red galaxies at all redshifts, and the $\tau$ models overestimate the apparent $R-I$ colors
of  $0.4\lesssim z \lesssim 0.7$ red galaxies by $\simeq 0.05$ magnitudes.
As discussed in \S\ref{sec:litcomp}, we expect these systematics to have a modest 
impact on our luminosity functions.

The $4^{\prime\prime}$ aperture photometry captures 86\% or less of the total flux. 
We corrected for the flux outside this aperture by assuming galaxies within our
sample have truncated \markcite{dev48}{de Vaucouleurs} (1948) profiles,
\begin{eqnarray}
I & \propto & {\rm exp}  \left[ -7.6695 \left(\frac{r}{r_e}\right)^{1/4} \right] \nonumber \\
  &         &  - {\rm exp}\left[ -7.6695 \times 7^{1/4} \right] 
\end{eqnarray}
at $r<7r_e$, where the half-light radius $r_e$ is a function of 
$B$-band absolute magnitude and redshift. At $z=0$, we use the size-luminosity relation of SDSS 
early-type galaxies \markcite{she03}({Shen} {et~al.} 2003) and assume $B_{\rm Vega}-r_{\rm AB}=1.32$ \markcite{fuk95}({Fukugita}, {Shimasaku}, \&  {Ichikawa} 1995), so 
\begin{equation}
{\rm log}[r_e({\rm h^{-1} kpc})]=1.0-0.26(M_B-5{\rm logh}+21.81).
\end{equation}
The $1\sigma$ dispersion around this relation is only $30\%$ 
for the most luminous red galaxies \markcite{she03}({Shen} {et~al.} 2003), while the offsets between 
$4^{\prime\prime}$ and total magnitudes in Figure~\ref{fig:sys} are a relatively weak
function of half-light radius. This simple relation should therefore provide accurate 
corrections between $4^{\prime\prime}$ aperture and total photometry.

\begin{figure*}
\epsscale{0.80}
\plotone{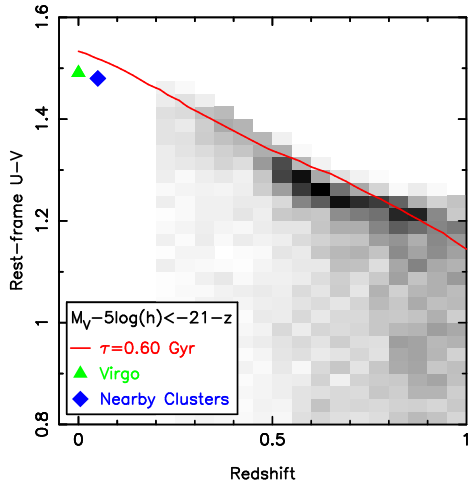}
\epsscale{1.0}
\caption{The evolving rest-frame $U-V$ colors of red and blue galaxies brighter than 
$M_V-5~{\rm log}~h=-21-z$. For comparison, the 
$U-V$ colors of $M_V-5~{\rm log}~h=-21$ red galaxies in Virgo \markcite{bow92}({Bower} {et~al.} 1992)
and $z=0.05$ clusters \markcite{mci05}({McIntosh} {et~al.} 2005) are also shown. 
The locus of galaxies which lie along the red galaxy color-magnitude relation
is evident. The evolution of the color-magnitude relation is well approximated by 
a \markcite{bru03}{Bruzual} \& {Charlot} (2003) $\tau=0.6~{\rm Gyr}$ model (solid red line), which has an exponentially 
declining star formation rate, no mergers, and little ongoing star formation at $z<1$.
\label{fig:uvz}}
\end{figure*}

\begin{figure*}
\vspace{1cm}
\epsscale{0.75}
\plotone{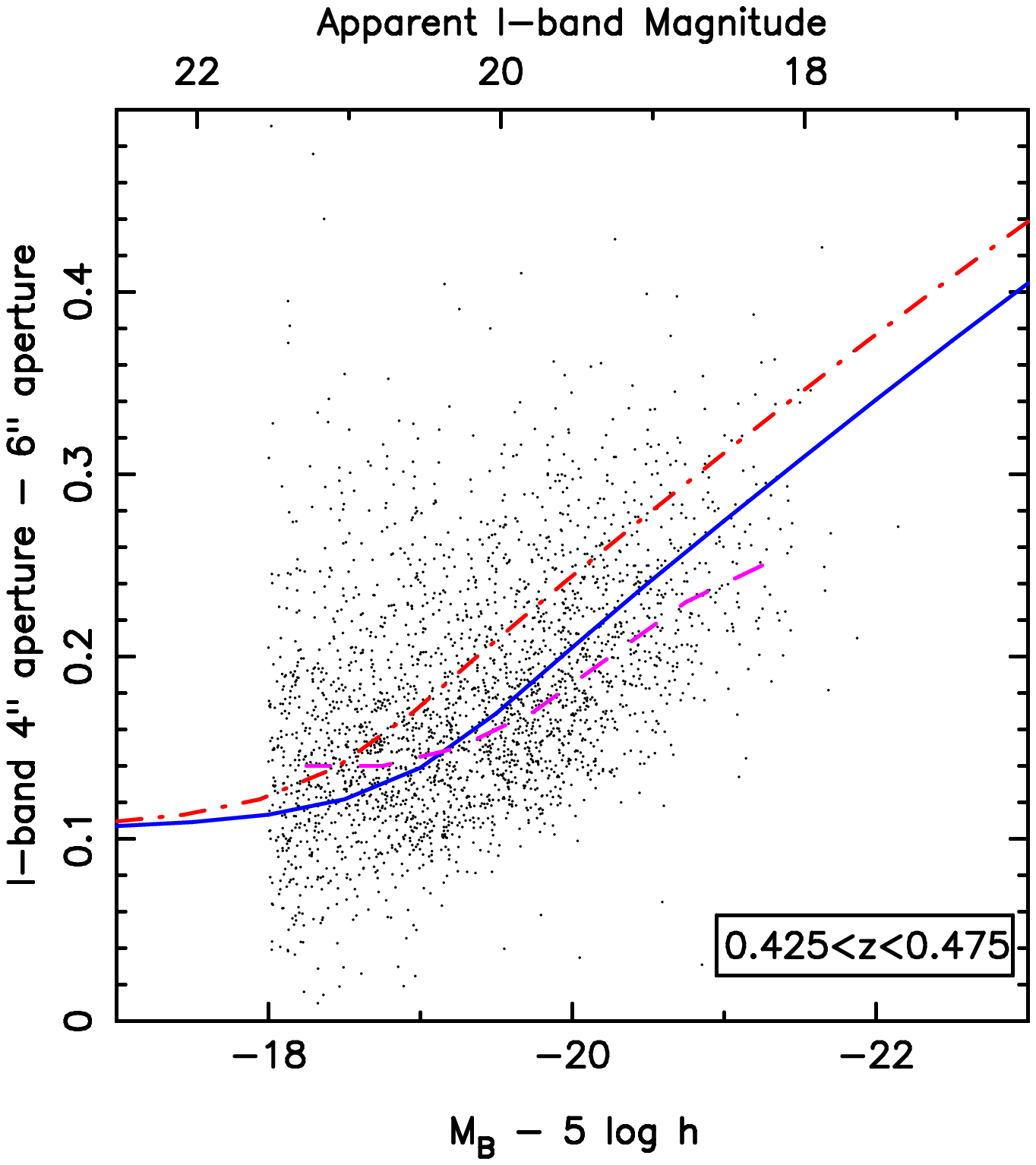}\plotone{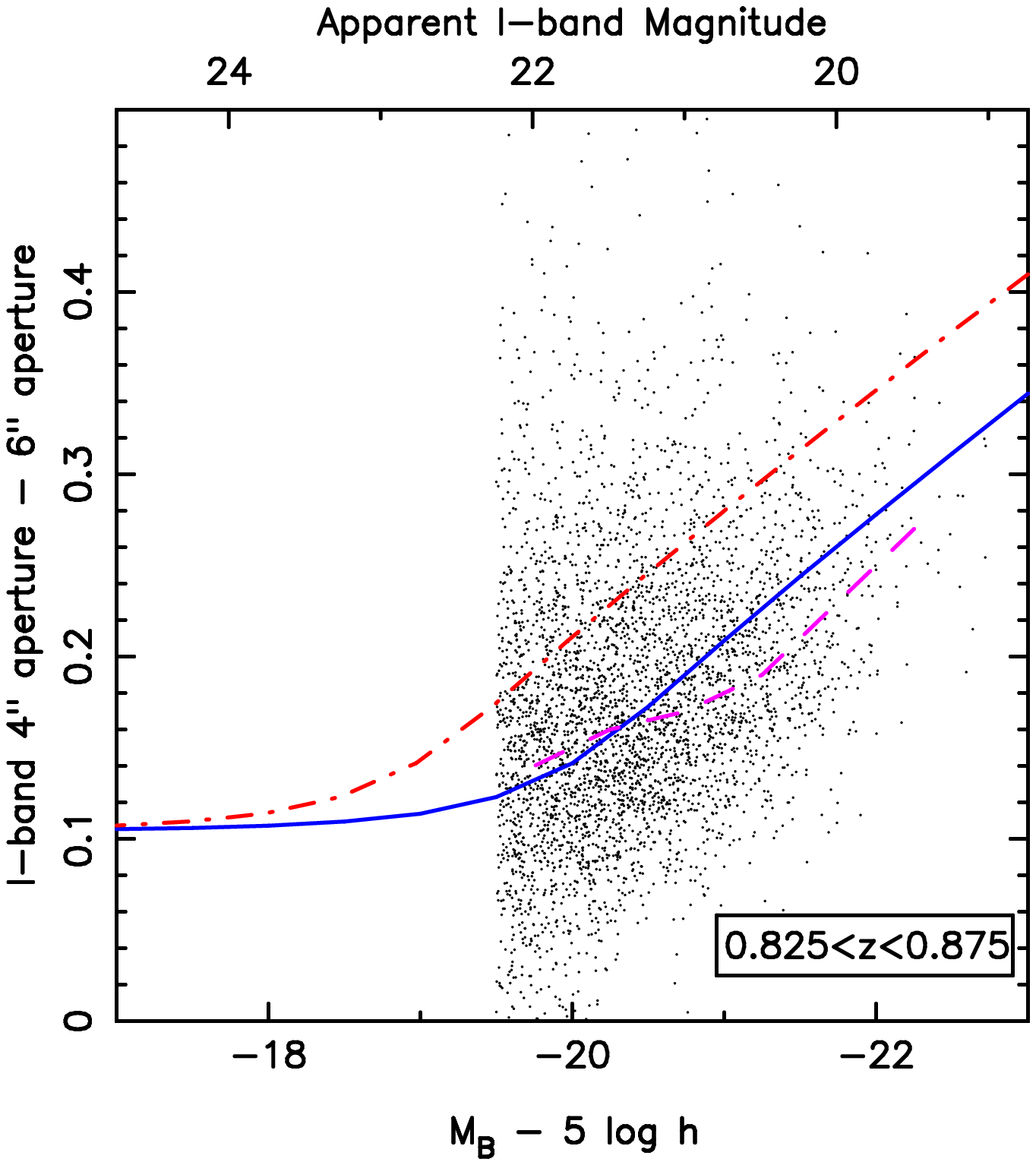}
\epsscale{1.00}
\caption{The observed and predicted offset between $4^{\prime\prime}$ and $6^{\prime\prime}$
aperture photometry for red galaxies. For the vast majority of $z>0.4$ red galaxies the measured 
offset is within $0.15$ magnitudes of the value expected for point sources ($0.11$ magnitudes). 
The median of the observed offsets is shown by a purple dashed line in both panels. 
The SDSS size-luminosity relation \markcite{she03}({Shen} {et~al.} 2003) without luminosity evolution (red dash-dot line) 
overestimates the sizes of red galaxies. The SDSS size-luminosity relation with luminosity 
evolution described by a $\tau=0.6~{\rm Gyr}$ \markcite{bru03}{Bruzual} \& {Charlot} (2003) model (blue solid line) approximates 
the data well. For a fixed {\it apparent} $I$-band magnitude, the observed half-light radius 
of red galaxies increases with increasing redshift. Estimators of total galaxy magnitudes may 
therefore exhibit larger systematic errors at $z=0.9$ than $z=0.5$. 
\label{fig:ap46}}
\end{figure*}

The size-luminosity relation must change with redshift due to the evolution of red 
galaxy stellar populations. We assume the size-luminosity relation undergoes pure 
luminosity evolution described by a \markcite{bru03}{Bruzual} \& {Charlot} (2003) $\tau=0.6~{\rm Gyr}$ stellar synthesis 
model with a formation redshift of $z=4$. This model fades by $1.24$ $B$-band magnitudes
between $z=1$ and $z=0$. As shown in Figure~\ref{fig:uvz}, the $\tau=0.6~{\rm Gyr}$ 
model approximates the color evolution of $M_V-5~{\rm log}~h \leq  -21-z$ red galaxies.
The luminosity evolution of the fundamental plane for $M_B-5~{\rm log}~h \simeq -20-1.2z$ 
early-type galaxies \markcite{van03,tre05}(e.g., {van Dokkum} \& {Stanford} 2003; {Treu} {et~al.} 2005) is also comparable to this model, though 
less luminous galaxies probably exhibit more rapid luminosity function \markcite{tre05,wel05}({Treu} {et~al.} 2005; {van der Wel} {et~al.} 2005).
Stellar population synthesis models with $\tau<1~{\rm Gyr}$ have less than $2\%$
of their total star formation at $z<1$. 
While this is a tiny amount of star formation, it does result in more rest-frame color 
and luminosity evolution than passive evolution models without any $z<1$ star formation.
Changing the value of $\tau$ by $0.2~{\rm Gyr}$ alters the model $z=1$
rest-frame $U-V$ colors and $B$-band luminosity evolution by $\simeq 0.1$ magnitudes.
After accounting for the evolving size-luminosity relation and our $1.35^{\prime\prime}$ point 
spread function, we find the $4^{\prime\prime}$ diameter aperture captures 75\% of the flux 
of a $M_B-5~{\rm log}~h=-21$ galaxy at $z=0.9$. 

Red galaxies do not exclusively have \markcite{dev48}{de Vaucouleurs} (1948) profiles \markcite{blan03,dri06}(e.g., {Blanton} {et~al.} 2003; {Driver} {et~al.} 2006), 
and this could affect our estimates of galaxy total magnitudes. To investigate this, we 
determined the  offsets between $4^{\prime\prime}$ aperture and total magnitudes for 
\markcite{ser68}{Sersic} (1968) profiles with indices between 2 and 6. For galaxies with half-light radii of 
$1^{\prime\prime}$, the offset differs from that determined with a \markcite{dev48}{de Vaucouleurs} (1948) profile by $0.03$ 
magnitudes. Changing the truncation radius from $7$ to $10$ half-light radii only changes 
the offset between $4^{\prime\prime}$ aperture and total magnitudes by $\sim 0.06$ magnitudes 
for galaxies with $r_e\sim 1^{\prime\prime}$. This offset can be increased by using
larger truncation radii, though this results in a systematic difference between 
the model and measured offset of $4^{\prime\prime}$ and $6^{\prime\prime}$
aperture photometry for red galaxies. As the expected evolution of red galaxy
stellar populations is on the order of $1.24$ $B$-band magnitudes per unit redshift, 
small errors in our corrections from $4^{\prime\prime}$ aperture to total magnitudes should
have little impact upon our results and conclusions.

To verify the accuracy of our model, we compared the measured and predicted 
difference between $4^{\prime\prime}$ and $6^{\prime\prime}$ aperture photometry 
for red galaxies. As shown in Figure~\ref{fig:ap46}, our simple model provides 
a good approximation of what is observed in our red galaxy sample. 
Figure~\ref{fig:ap46} also illustrates the absolute magnitude dependence of our model. 
This dependence results in the aperture corrections at fixed {\it apparent}
$I$-band magnitude increasing with redshift. It is therefore plausible that 
an estimator of total magnitudes could work well for the majority of $I=21$ galaxies, which are
at $z<0.7$, but have systematic errors for $I=21$ galaxies at $z=0.9$. 
Accounting for flux outside the aperture is crucial for measuring the total
magnitudes of $z<1$ galaxies.

\section{THE RED GALAXY SAMPLE}
\label{sec:sample}

The rest-frame distribution of galaxy colors is bimodal \markcite{hog04}(e.g., {Hogg} {et~al.} 2004), and selection
criteria for red galaxies typically fall near the minimum between 
the red and blue galaxy populations \markcite{mad02,bel04,fab05}(e.g., {Madgwick} {et~al.} 2002; {Bell} {et~al.} 2004; {Faber} {et~al.} 2005). We apply a similar 
approach for this work, and use the following rest-frame color selection 
criterion;
\begin{eqnarray}
\nonumber \\
U-V & > & 1.40-0.25 \nonumber \\
    &   & -0.08\times(M_V-5~{\rm log}~h+20.0) \nonumber \\
    &   & -0.42\times(z-0.05) \nonumber \\
    &   & +0.07\times(z-0.05)^2.
\end{eqnarray}
Our criterion selects galaxies with rest-frame $U-V$ colors within 
$0.25$ magnitudes of the evolving color-magnitude relation of red galaxies.
This criterion allows comparison with the recent literature and 
is very similar, though not identical, to the criterion of \markcite{bel04}{Bell} {et~al.} (2004).

\begin{figure}[t]
\plotone{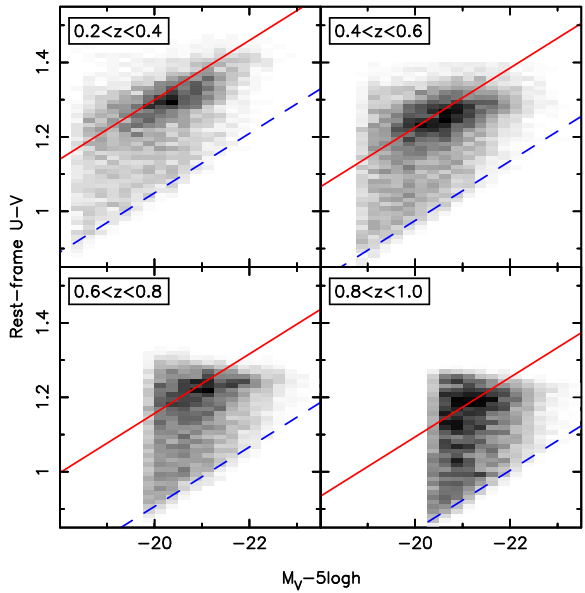}
\caption{Color-magnitude diagrams for red galaxies.
Our selection criterion is shown with the dashed blue line and is 
0.25 magnitudes blueward of the model color-magnitude relation (solid red line). The 
color-magnitude relation moves towards bluer rest-frame colors with increasing 
redshift. There is a difference between the model and observed color-magnitude relation
slope at $0.2<z<0.4$, due to our use of $4^{\prime\prime}$ aperture photometry, which 
measures the cores of low redshift galaxies. As discussed in \S\ref{sec:rest}, this has 
negligible impact on our luminosity functions and conclusions.
\label{fig:cmr}}
\end{figure}

Our selection criterion is plotted in Figure~\ref{fig:cmr} along with the
observed color distribution of red galaxies. Our selection criterion
has slightly more tilt than the observed color-magnitude relation of $0.2<z<0.4$
red galaxies. If we use $8^{\prime\prime}$ aperture photometry instead of $4^{\prime\prime}$ 
aperture photometry, the observed color-magnitude relation of $0.2<z<0.4$ red galaxies 
moves blueward and steepens. However, as this has little impact upon 
the measured luminosity function, we continue to use $4^{\prime\prime}$ aperture 
photometry, which has smaller random uncertainties than $8^{\prime\prime}$ aperture photometry.

The measured space density of the most luminous red galaxies does not depend
on the details of our selection criterion, as these galaxies mostly lie
along the color-magnitude relation. We can therefore easily compare our 
luminosity functions for $M_B-5~{\rm log}~h<-21$ red galaxies with those of 
2dFGRS, SDSS, COMBO-17, and DEEP2 \markcite{mad02,bel04,fab05,wil05,bla05}({Madgwick} {et~al.} 2002; {Bell} {et~al.} 2004; {Faber} {et~al.} 2005; {Willmer} {et~al.} 2005; {Blanton} 2005), 
all of which use slightly different galaxy selection criteria.
The fraction of blue galaxies increases with decreasing luminosity, 
so the measured space density of $\lesssim L^*$ red galaxies is a stronger function
of the red galaxy selection criteria. If we shift our criterion blueward by 
$0.10$ magnitudes our measured space density of  $M_B-5~{\rm log}~h\sim -20$ red galaxies
increases by $\simeq 25\%$.

\begin{figure}[hbt]
\plotone{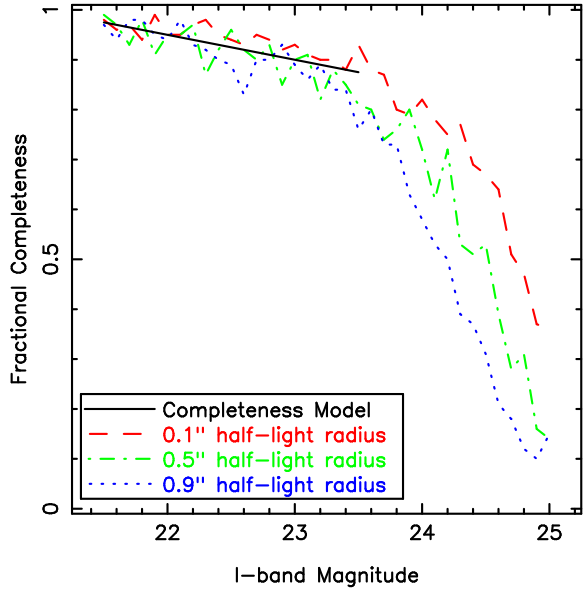}
\caption{The completeness of the red galaxy sample as a function of
$I$-band apparent magnitude. We measured the completeness by adding
artificial galaxies to copies of the data and recovering them with 
SExtractor. As our incompleteness is dominated by confusion with 
brighter (and generally lower redshift) sources rather than fluctuations
in the sky background, our completeness is a weak function of half-light radius.
The completeness is higher than 85\% over our sample magnitude range, and is 
well approximated by $1-0.05(I-21)$ for $21.0<I<23.5$ galaxies.
\label{fig:comp}}
\end{figure}

We limit the absolute magnitude range in each of our redshift bins 
so we can determine accurate redshifts and have a highly complete sample. 
Almost all of the galaxies in our final sample are detected in 
$B_W$, $R$, $I$ and the two IRAC bands. To evaluate the completeness 
of our sample, we added artificial galaxies to copies of the $I$-band images 
and recovered them with SExtractor. The artificial galaxies had \markcite{dev48}{de Vaucouleurs} (1948) 
profiles truncated at 7 effective radii, and we measure the completeness
for a range of effective radii and apparent magnitudes.
As illustrated in Figure~\ref{fig:comp}, our catalogs are more than 
85\% complete for $I<23.5$ galaxies with half-light radii of $0.5^{\prime\prime}$ or less. 
Our measurements of the luminosity function, which we discuss in \S\ref{sec:lf}, include 
small corrections for this incompleteness.

\subsection{ADDITIONAL SELECTION CRITERIA}

\begin{figure*}[hbt]
\vspace{1cm}
\epsscale{1.00}
\plotone{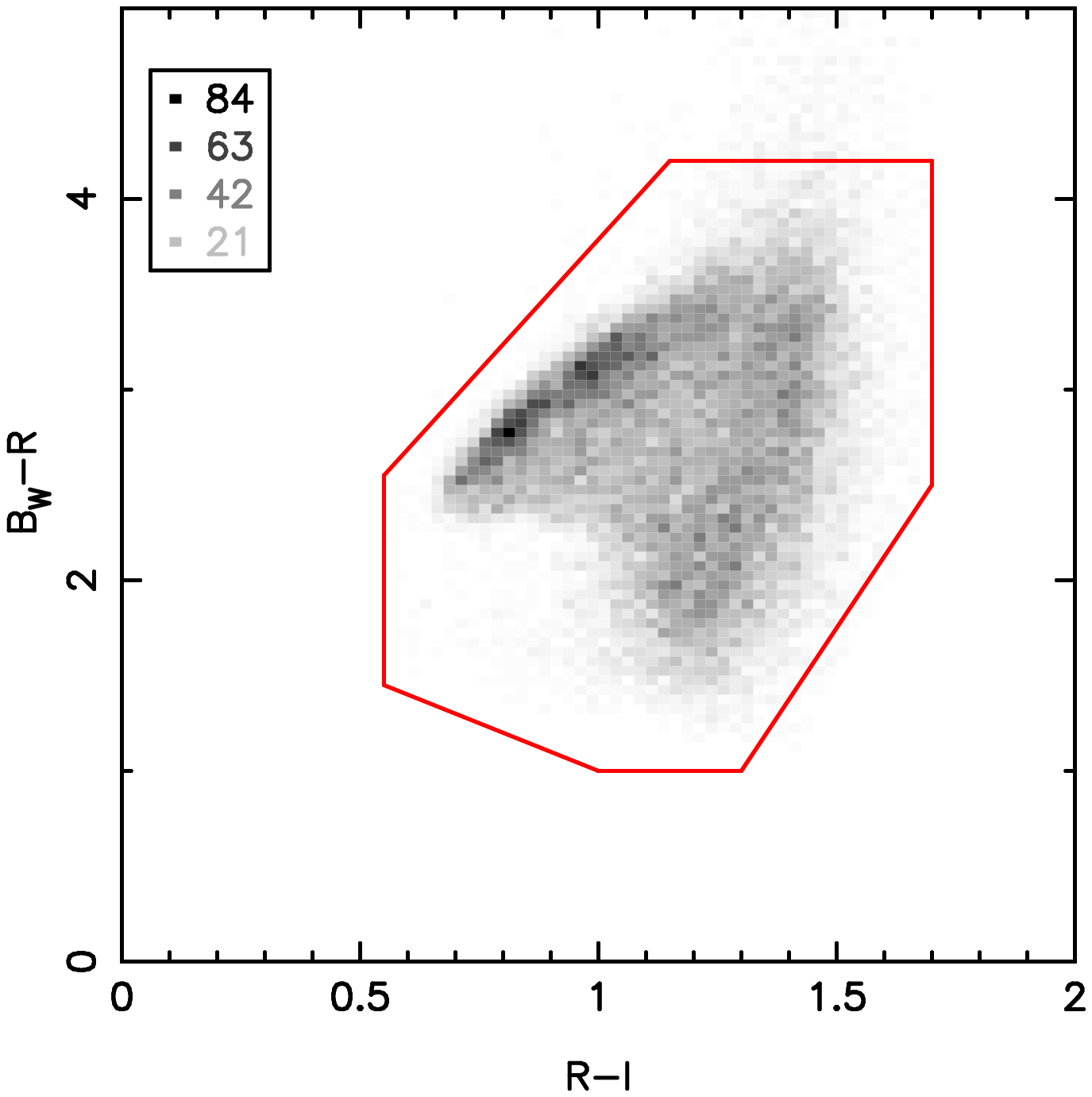}\epsscale{0.98}\plotone{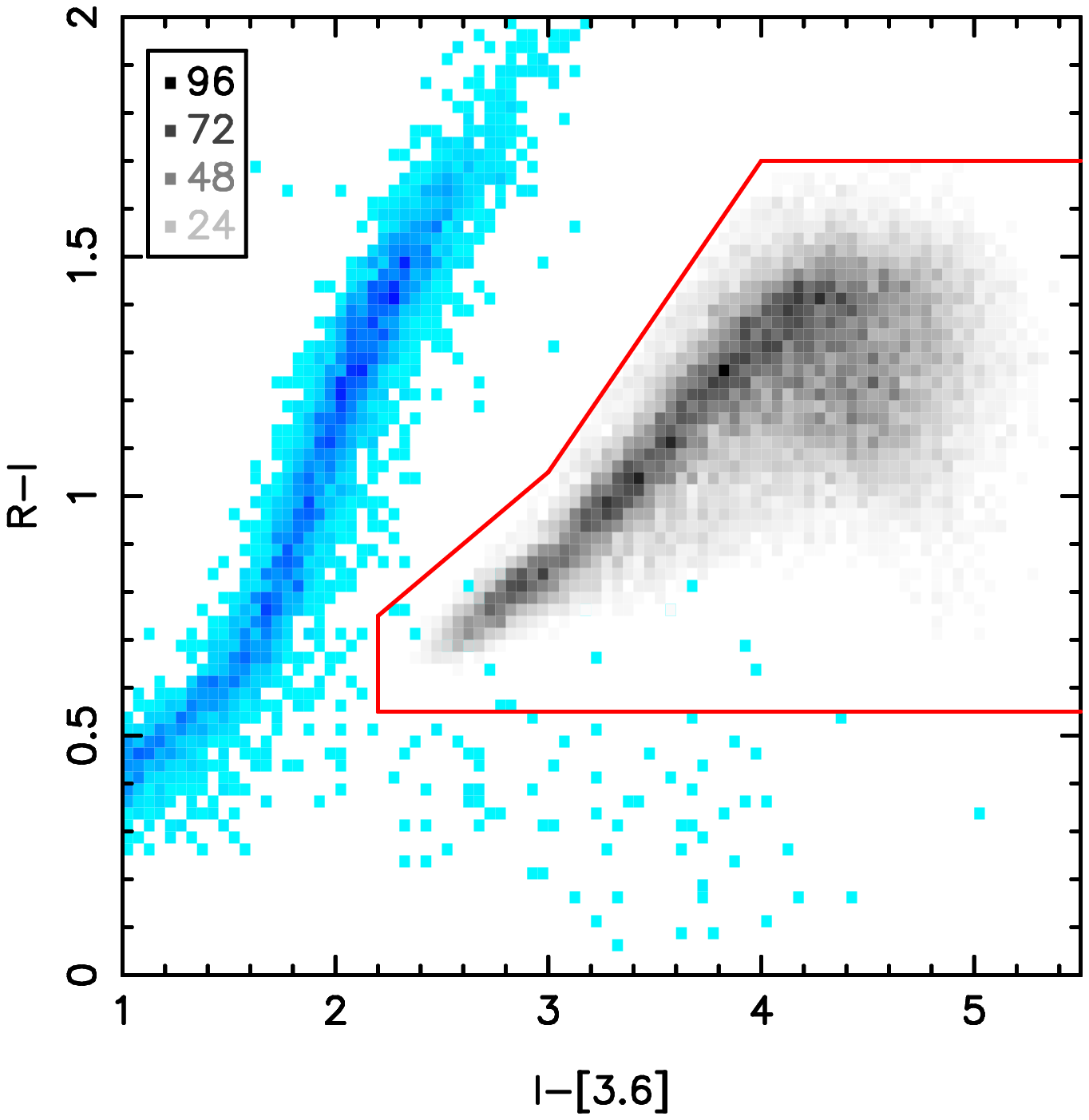}
\epsscale{1.00}
\caption{The apparent color distributions of $M_B-5{\rm log}h<-19.5$ red galaxies in \bootes.
Red lines denote apparent color cuts, which we use to exclude stars, 
quasars, blue galaxies, $z \gg 1$ galaxies, and gross photometric outliers.
For faint objects with highly uncertain apparent colors (e.g., $B_W>26.5$ galaxies), 
we use upper or lower limits  for the color when applying the color cuts.
The locus of $18<I<19$ stars and quasars is shown in blue in the right-hand
panel, and these objects are largely excluded by our color cuts.
The red galaxy color-magnitude relation produces an over-density in color space that is 
clearly evident.
\label{fig:bwri}}
\end{figure*}

Contaminants cannot be rejected from photometric redshift surveys on the 
basis of their spectra, so these surveys can be more susceptible to 
contaminants than comparable spectroscopic surveys. 
This contamination can include stars, quasars, and galaxies with 
large photometric redshift errors. We therefore apply apparent color and 
morphology cuts to minimize contamination while retaining a highly
complete sample of red galaxies.

We have applied apparent color cuts to exclude objects whose
colors differ from those expected for $0.2<z<1.0$ red galaxies.
These color cuts are designed to remove most stars, quasars and $z\gg1$
galaxies from our sample.  Our cuts may exclude a small percentage of red 
galaxies with unusually large photometric errors, but as these galaxies
could also have large redshift and luminosity errors, it is preferable to
exclude them from our measurement of the luminosity function.
As the distribution of red galaxy apparent colors does not form
a simple shape in color space, we use a total 17 color cuts to exclude 
objects from the sample. 
All but two of these cuts are plotted in Figure~\ref{fig:bwri},
while a full list is provided in Table~\ref{table:colcut}.

Our color cuts remove 10767 objects from the sample, reducing it to 
39866 objects. As we apply the color cuts before applying the morphology 
criterion, most of the objects rejected from the sample have the colors
and morphologies of stars and quasars.
Unlike $0.2<z<1.0$ red galaxies, main sequence stars have 
apparent colors which satisfy $R-I>0.65\times(I-[3.6]-1)$ or $I-[3.6]<2.2$, 
and 8775 of the objects excluded by our color cuts satisfy these two criteria. 
Another 749 of the excluded objects satisfy $[3.6]-[4.5]>0.6$, and are probably
quasars. The measured photometry of the remaining 1223 objects excluded by our color cuts
differs somewhat from main sequence stars, most quasars, and $0.2<z<1.0$ red galaxies.
Spectroscopic redshifts are available for 23 of these objects, and 13 are $0.2<z<1.0$
galaxies while the remainder are stars and quasars. It is therefore plausible that 
$\sim 700$ red galaxies are excluded by our color cuts, though this is only
$2\%$ of our final sample. 

To further reduce contamination by compact objects, we also apply a morphology 
criterion. We exclude objects if the difference between their $2^{\prime\prime}$ and 
$4^{\prime\prime}$ $I$-band aperture photometry is 0.20 magnitudes below the 
expectation from the size-luminosity relation discussed in \S\ref{sec:rest}.
We expect many of the 287 objects rejected by this criterion to be blends of 
galaxies with stars and quasars. The morphology and apparent color cuts
could plausibly exclude $\simeq 10^3$ red galaxies from our final sample. 
This reduces sample size by $\simeq 2.5\%$, but as our uncertainties from 
cosmic variance are on the order of $10\%$ (\S\ref{sec:lf}), the impact upon our results and 
conclusions is insignificant.
The final $0.2<z<1.0$ red galaxy sample contains 39599 galaxies brighter 
than $I=23.5$, and number counts as a function of both redshift and apparent 
magnitude are summarized in Table~\ref{table:counts}.

\section{THE RED GALAXY LUMINOSITY FUNCTION}
\label{sec:lf}

We measured the red galaxy luminosity function in four redshift slices 
between $z=0.2$ and $z=1.0$. We used both the non-parametric $1/V_{max}$ 
technique \markcite{sch68}({Schmidt} 1968) and maximum likelihood fits \markcite{mar83}(e.g., {Marshall} {et~al.} 1983) of 
\markcite{sch76}{Schechter} (1976) functions;
\begin{equation}
\phi(M) {\rm d} M
= 0.4 {\rm ln} 10 \times \phi^* \left(\frac{L}{L^*}\right)^{\alpha+1} {\rm exp} \left(\frac{-L}{L^*}\right) {\rm d} M, 
\end{equation}
where $L$ is the galaxy luminosity while $\phi^*$, $L^*$, and $\alpha$ are constants. 
Like most luminosity function papers, we use $M^*$ rather than $L^*$, where $M-M^*=-2.5{\rm log} (L/L^*)$.

The uncertainties of the luminosity function are dominated by 
large-scale structure rather than Poisson counting statistics. 
As galaxies with different luminosities can occur within the same 
large-scale structures, the data-points in binned luminosity functions 
are not independent of each other. We have evaluated the uncertainties 
of the luminosity function using both subsamples of the \bootes~field and the 
galaxy angular correlation function.

\begin{figure}[hbt]
\plotone{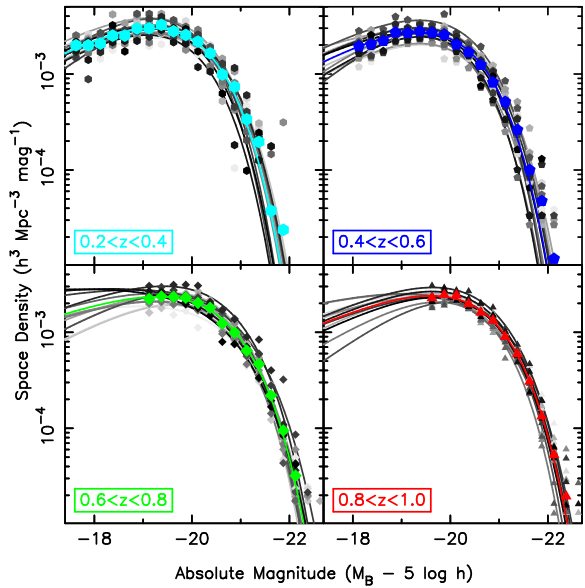}
\caption{Red galaxy luminosity functions for the entire \bootes~field and our thirteen subsamples.
The solid lines are maximum likelihood Schechter function fits to the data while the 
symbols are $1/V_{max}$ estimates of the luminosity function. 
The luminosity functions for each $0.5~{\rm deg}^2$ subsample are shown in a different greyscale while 
the luminosity functions for the entire \bootes~field are shown in color. 
While individual galaxy clusters are evident in Figure~\ref{fig:sky}, these contain 
only a fraction of all red galaxies and the variations between different subsamples are almost 
certainly caused by galaxies residing within larger structures.
\label{fig:lf_sub}}
\end{figure}

\begin{figure*}[t]
\epsscale{1.30}
\plotone{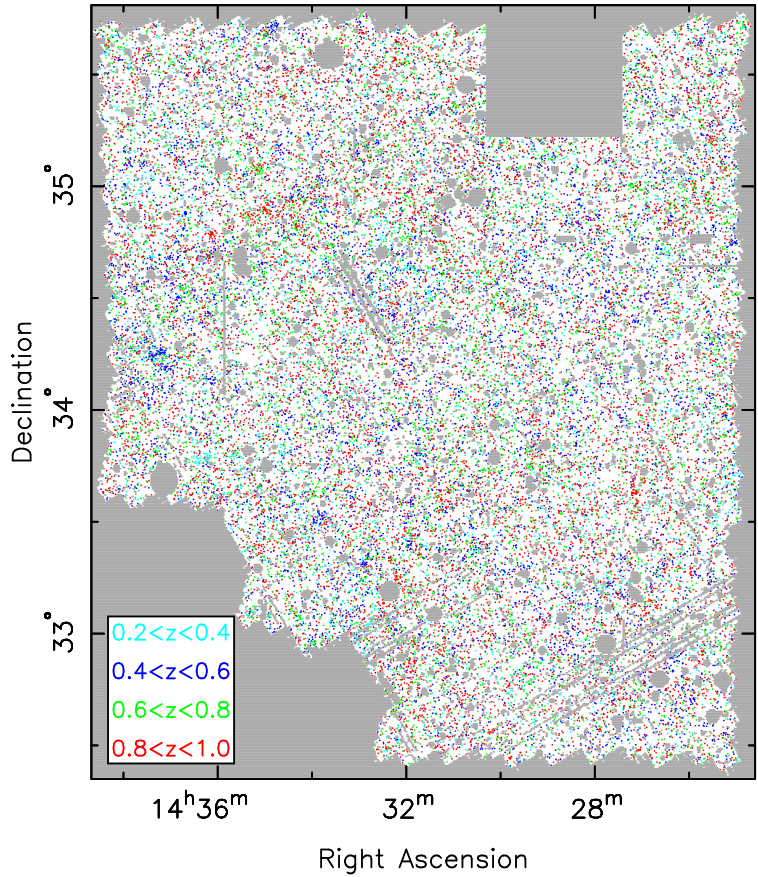}
\epsscale{1.00}
\caption{The sky distribution of red galaxies in the $2.7^\circ \times 3.3^\circ$ Bootes field.
In the electronic edition, $0.2<z<0.4$, $0.4<z<0.6$,  $0.6<z<0.8$ and $0.8<z<1.0$ objects are denoted with 
light blue, dark blue, green and red symbols respectively. Grey regions denote areas
excluded from the sample, including saturated stars and areas without both NDWFS and {\it Spitzer} coverage.
Individual structures are clearly evident, and $0.4<z<0.6$ galaxy clusters are not uniformly 
distributed across the \bootes~field.
At $z>0.6$, the distribution of galaxies is more uniform, though individual structures 
can still be seen. Of particular note is a $\sim 30 h^{-1} {\rm Mpc}$ long $z\simeq 0.93$ 
structure at $14^{\rm h}35^{\rm m}30^{\rm s},~+34^{\circ}50^{\prime}$.
\label{fig:sky}}
\end{figure*}

Subsamples are conceptually simple but underestimate the uncertainties, 
as an individual large-scale structure may span 2 subsamples of the data. 
For this reason, we only use thirteen $0.5~{\rm deg}^2$ subsamples rather
than many smaller subsamples. For our  Schechter function fits we evaluate 
the luminosity function for each subsample using the  method of \markcite{mar83}{Marshall} {et~al.} (1983) 
and use the standard deviation of the fitted parameters (e.g., $M^*$) divided 
by $\sqrt{13}$ to estimate uncertainties. Luminosity functions for the 
whole \bootes~field and the thirteen subsamples are shown in Figure~\ref{fig:lf_sub}. 
The subample luminosity functions can differ from the luminosity function for the
entire field by as much as $50\%$. 
While individual galaxy clusters are evident in Figure~\ref{fig:sky}, these contain 
but a fraction of all red galaxies and the variations between different subsamples 
are almost certainly caused by galaxies residing within larger structures.

As the angular correlation function of galaxies does not equal zero on scales 
of $\sim 1^{\circ}$, we expect subsamples to underestimate the uncertainties for
$\phi^*$ and the luminosity density, $j_B$. We do not calculate uncertainties
for $M^*$ and $\alpha$ using galaxy clustering, as this requires additional information 
including details of how the shape of the luminosity function varies with galaxy density.
The expected variance of the number counts in a field is given by
\begin{equation}
\left< \frac{n_i-<n_i>}{n_i}\right>^2=\frac{1}{<n_i>}+\frac{1}{\Omega^2} \int \int \omega(\theta) d\Omega_1 d\Omega_2
\end{equation}
\markcite{gro77,efs91}({Groth} \& {Peebles} 1977; {Efstathiou} {et~al.} 1991) where $\omega(\theta)$ is the angular correlation function, $\theta$ is the angle separating
solid angle elements $d\Omega_1$ and $d\Omega_2$, and $\Omega$ is the area of the field.
We assume $\omega(\theta)$ is a power-law with index $1-\gamma$, and use power-law fits to the angular 
correlation functions from M.~J.~I. Brown et~al. (in preparation). 
For the \bootes~field the sample variance is approximately $n_i^2 \omega(1^\prime) 10^{-1.78(\gamma-1)}$,
while for fields smaller than $1~{\rm deg^2}$ the sample variance is at least $n_i^2 \omega(1^\prime) 10^{-1.33(\gamma-1)}$.
Table~\ref{table:clu} lists uncertainties for $\phi^*$ and $j_B$ derived from subsamples
and the angular clustering, along with the $\omega(1^{\prime})$ and $\gamma$ values from M.~J.~I. Brown et~al. (in preparation). 
Subsamples underestimate the uncertainties of $\phi^*$ and $j_B$ by a factor of $\simeq 3$ at
low redshift, so throughout the remainder of the paper we use angular clustering uncertainties for  $\phi^*$ and $j_B$.

\begin{figure*}[hbt]
\plotone{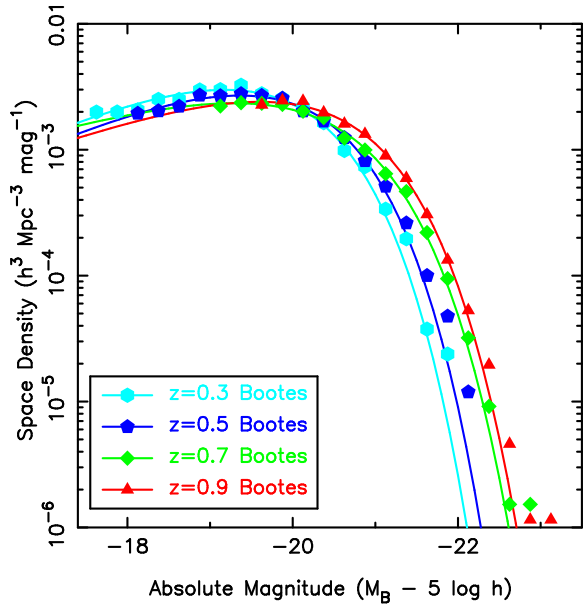}
\caption{Luminosity functions for our red galaxy sample. 
The solid lines are maximum likelihood Schechter function fits to the data while the 
symbols are $1/V_{max}$ estimates of the luminosity function. We use the same symbols
for our four redshift bins throughout the remainder of the paper.  
For clarity, we do not plot the $1\sigma$ uncertainties here but list them in Table~\ref{table:vmax}.
For most of our redshift and luminosity bins, these uncertainties are on the order of $10\%$. 
Within our sample, red galaxies brighter than $L^*$ are clearly evolving, while $\alpha$ and $\phi^*$ do not
show a strong trend with increasing redshift.
\label{fig:lfz}}
\end{figure*}

Our $1/V_{max}$ luminosity function values and Schechter function fit 
parameters are provided in Tables~\ref{table:vmax} and~\ref{table:mlf} 
respectively. For comparison to the prior literature, fits with 
$\alpha=-0.5$ are also provided in Table~\ref{table:mlf} though 
for the remainder of the paper we use fits where $\alpha$ is a free parameter. 
As shown in Figure~\ref{fig:lfz}, Schechter functions 
provide a good approximation of the observed luminosity function.

It is evident from the fits shown in Figure~\ref{fig:lfz} that the red galaxy luminosity function 
evolves with redshift. In particular, the bright end of the luminosity function (i.e., at $L\gtrsim L^*$)
steadily fades by $\simeq 0.9$ magnitudes per unit redshift from $z=0.9$ to $z=0.3$. 
While $\alpha$ may decrease slightly with increasing redshift, the fits computed for each 
redshift are constrained by different ranges of galaxy luminosity and the correlation of 
$\alpha$ with redshift disappears when we only fit the luminosity function of 
$M_B-5{\rm log} h<-19-0.9z$ red galaxies. From Figure~\ref{fig:lfz} and Table~\ref{table:vmax} 
it is unclear if the luminosity function is or is not undergoing density evolution.
While an apparent decline in the space density is measured (see Table~\ref{table:vmax}), the 
$1/V_{\rm max}$ luminosity function bins are correlated with each other and Figure~\ref{fig:lfz} 
does not include the uncertainties from cosmic variance, which are  $\simeq 0.1\phi(M)$.
The normalization of the luminosity function, $\phi^*$, also does not show a consistent trend with 
redshift. We note, however, that $\phi^*$ is very sensitive to the measured and assumed values of 
both $M^*$ and $\alpha$.

In \S\ref{sec:jbsec} and \S\ref{sec:msec} we measure the assembly
and evolution of red galaxies using the luminosity density and 
the luminosity evolution of galaxies at a fixed space density 
threshold. These parameterizations of the evolving luminosity
function are not highly correlated with Schechter function parameters 
such as $\alpha$. Also, these parameters can be easily compared with 
simple models of red galaxy assembly and stellar population evolution, thus simplifying
the interpretation of the evolving luminosity function of red galaxies.

\subsection{THE LUMINOSITY DENSITY AND THE EVOLVING STELLAR MASS WITHIN RED GALAXIES}
\label{sec:jbsec}

\begin{figure*}[hbt]
\epsscale{1.00}
\plotone{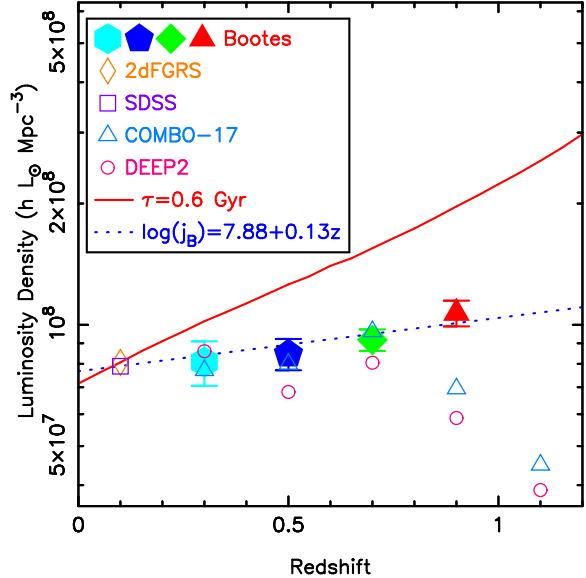}
\epsscale{1.00}
\caption{The $B$-band luminosity density of red galaxies with redshift.
While a \markcite{bru03}{Bruzual} \& {Charlot} (2003) $\tau=0.6~{\rm Gyr}$ model can approximate the colors
of red galaxies, it overestimates the evolution of the luminosity density. 
A fit to our data and the 2dFGRS is shown with the dashed blue line.
This fit indicates that the luminosity density has increased by $36\pm13\%$ from $z=0$ to $z=1$, 
rather than the 213\% predicted by the $\tau=0.6~{\rm Gyr}$ model.
If red galaxy stellar populations have faded by $\simeq 1.24$ $B$-band magnitudes since $z=1$,
the stellar mass contained within the ensemble of red galaxies has approximately doubled over the same period.
\label{fig:jb}}
\end{figure*}

We use the luminosity density to measure the rest-frame $B$-band light emitted by ensembles of 
red galaxies. The luminosity density is strongly correlated with the total stellar 
mass contained within all red galaxies, though it is also sensitive to the evolving 
stellar populations of these galaxies. To evaluate the luminosity density 
for all luminosities, we integrate over the best-fit Schechter functions, so 
\begin{equation}
j_B=\phi_B^*L_B^*\Gamma(\alpha+2).
\end{equation}
We present the luminosity density of all red galaxies as function of
redshift in Figure~\ref{fig:jb}. Luminosity density values are listed
in Table~\ref{table:jbtab}, including values determined with galaxies
brighter than evolving absolute magnitude thresholds which are a function of $M^*$. 
For our range of $\alpha$ values, less than 25\% of the luminosity density
is contributed by galaxies fainter than our magnitude limits and approximately 70\% of 
the luminosity density is contributed by galaxies within a magnitude of $M^*$.
Our measurements are broadly consistent with the prior literature, except
at $z=0.9$ where we find a higher luminosity density than COMBO-17 and DEEP2.

To model the evolution of the luminosity density, we assume that the stellar 
populations of red galaxies can be approximated by a \markcite{bru03}{Bruzual} \& {Charlot} (2003) $\tau=0.6~{\rm Gyr}$ 
model with a formation redshift of $z=4$. This model approximates the evolving 
rest-frame colors of $M_V-5{\rm log}h \sim -21$ red galaxies, which we plot in Figure~\ref{fig:uvz}.
The $\tau=0.6~{\rm Gyr}$ model fades by $1.24$ $B$-band magnitudes between $z=1$ and $z=0$, 
which is consistent with the observed luminosity evolution of both the size-luminosity 
(Figure~\ref{fig:ap46}) and fundamental plane relations \markcite{van03,tre05}(e.g., {van Dokkum} \& {Stanford} 2003; {Treu} {et~al.} 2005) 
of $M_V-5{\rm log}h \sim -21$ red galaxies. We caution that less luminous red galaxies are 
bluer than the $\tau=0.6~{\rm Gyr}$ model and their stellar populations may exhibit more rapid 
luminosity evolution at $z<1$ \markcite{tre05,wel05}(e.g., {Treu} {et~al.} 2005; {van der Wel} {et~al.} 2005). 
The $\tau=0.6~{\rm Gyr}$ model, normalized to the 2dFGRS, is plotted in Figure~\ref{fig:jb} 
and grossly overestimates the evolution of the luminosity density of red galaxies.

The relationship between ${\rm log}(j_B)$ and redshift in Figure~\ref{fig:jb} 
can be approximated by a straight line. If we fit to the \bootes~data alone, we find
\begin{eqnarray}
{\rm log }[j_B (10^{7} h~L_{\odot}~{\rm Mpc^{-3}})] & = & 7.82(\pm 0.06) \nonumber \\
                                                    &   &  + 0.22(\pm 0.09) \times z, \nonumber \\
\end{eqnarray}
with a reduced $\chi^2$ of only $0.2$. While the $\tau=0.6~{\rm Gyr}$ model predicts a $213\%$
increase in the $B$-band luminosity density of red galaxies between $z=0$ and $z=1$, using the 
\bootes~data alone we observe an increase of only $65\pm 34\%$.

Although we have the largest $0.2<z<1.0$ red galaxy sample currently available, our 
measurement of $j_B$ evolution is not sufficiently precise to clearly distinguish between significantly 
different descriptions of luminosity function evolution: pure luminosity evolution (where $j_B$ 
evolves in the same manner as $L^*$) or luminosity evolution with up to 50\% density evolution 
between $z=0$ and $z=1$. While combining data sets can introduce systematic errors, the 
combined data set allows the investigation of $j_B$ over a longer redshift baseline and reduces 
the random uncertainties at low redshift. The 2dFGRS red galaxy sample is selected using 
a criterion based upon principal component analysis of 2dF spectra \markcite{mad02}({Madgwick} {et~al.} 2002), which is
strongly correlated with galaxy rest-frame color. By fitting \markcite{bru03}{Bruzual} \& {Charlot} (2003) $\tau$ models
to galaxies with SDSS photometry and 2dFGRS spectroscopy, we find the 2dF red galaxy selection 
criterion corresponds to rest-frame $U-V\gtrsim 1.1$, which is very similar to our criterion for 
$L^*$ red galaxies at low redshift. Using the combined data set, we find
\begin{eqnarray}
{\rm log }[j_B (10^{7} h~L_{\odot}~{\rm Mpc^{-3}})] & = & 7.88(\pm 0.02) \nonumber \\ 
                                                    &   & + 0.13(\pm 0.04) \times z, \nonumber \\
\end{eqnarray}
which has a reduced $\chi^2$ of $0.5$. Our best fit to the combined data set, shown in 
Figure~\ref{fig:jb}, yields a $36\pm13\%$ increase in the luminosity density of red galaxies 
between $z=0$ and $z=1$.

We can infer the rate of stellar mass growth within the red galaxy population by comparing the 
evolution of $j_B$ with the luminosity evolution predicted by a stellar population synthesis model.
We assume red galaxy stellar populations have faded by $1.24$ $B$-band magnitudes since $z=1$, as 
does the $\tau=0.6~{\rm Gyr}$ model. We caution that if we underestimate the luminosity evolution 
of the stellar populations, we underestimate the growth of stellar mass within red galaxies, and vice 
versa. If red galaxy stellar populations fade by $1.24$ $B$-band magnitudes per unit redshift, 
then using the \bootes~data alone we find the stellar mass contained within the red galaxy 
population has increased by $91\pm 39\%$ since $z=1$. Using both \bootes~and the 2dFGRS, we
find the stellar mass contained within the red galaxy population has increased by $131\pm 22\%$
since $z=1$. Unless we have grossly overestimated the luminosity evolution of red galaxy
stellar populations, the stellar mass contained within the red galaxy population has increased
by order unity since $z=1$.


The mild evolution of red galaxy luminosity density rules out several models of red 
galaxy evolution. Stellar population synthesis models with roughly constant star formation 
histories exhibit modest evolution of their $B$-band luminosities at $z<1$.
However, such star formation histories result in galaxies with rest-frame 
colors of $U-V<0.5$, which is bluer than any of our red galaxies. While red 
galaxy mergers occur at $z<1$ \markcite{lau88,van05}(e.g., {Lauer} 1988; {van Dokkum} 2005), unless they are accompanied by star 
formation they will only redistribute stellar mass already within the red galaxy population. 
Stellar mass must be added to the red galaxy population from an outside 
source, namely blue galaxies.

Blue galaxies must be adding mass to the red population due to a 
decline in their star formation rates. Stellar population synthesis 
models with $1.5~{\rm Gyr}<\tau<2.5~{\rm Gyr}$ move across our 
rest-frame $U-V$ selection criterion for $L^*$ galaxies between 
$z=1$ and $z=0$. Stellar population synthesis models can also 
move across our selection criterion if star formation is rapidly 
truncated, possibly after AGN feedback or a merger triggered starburst. 
While a broad range of star formation histories can transform blue 
galaxies into red galaxies, the tight color-magnitude relation of red galaxies 
indicates that red galaxies of a given mass and metallicity have similar low rates of 
star formation. These galaxies have similar star formation histories
over many ${\rm Gyr}$, or they converge towards similar yet low 
star formation rates upon entering the red galaxy population.
We will explore some of these possibilities in detail in future papers.

\subsection{THE EVOLUTION OF VERY LUMINOUS RED GALAXIES}
\label{sec:msec}

Fading star-forming galaxies cannot explain the evolution of 
$4L^*$ red galaxies at $z<1$, as star-forming galaxies 
with comparable masses are exceptionally rare at these 
redshifts \markcite{bel04}(e.g., {Bell} {et~al.} 2004). If the stellar mass 
contained within these very luminous red galaxies is 
evolving at $z<1$, this is presumably due to galaxy mergers.

The luminosity density is a poor measure of the evolution of $4L^*$ red galaxies. 
As these galaxies are on the exponential part of the luminosity function, the 
measured luminosity density and space density of galaxies brighter than 
an absolute magnitude threshold will have an extremely strong dependence upon 
that threshold. For example, in Table~\ref{table:jbtab} we list values of 
$j_B$ for $M_B<M^*-1.5$ galaxies, but $M^*$ is correlated with $\alpha$ so 
the $j_B$ values determined with floating and fixed values of $\alpha$ differ by up to $35\%$.
Similarly, if the measured luminosities or the assumed luminosity evolution of $4L^*$ red
galaxies are in error 0.1 magnitudes, the measured space density evolution of 
red galaxies can be in error by 30\%. If the stellar populations of today's $L^*$ and 
$4L^*$ galaxies did not evolve in the same manner, luminosity density values derived 
with $M_B<M^*-1.5$ red galaxies can provide a misleading picture of $4L^*$ red galaxy evolution.

\begin{figure*}[hbt]
\epsscale{1.00}
\plotone{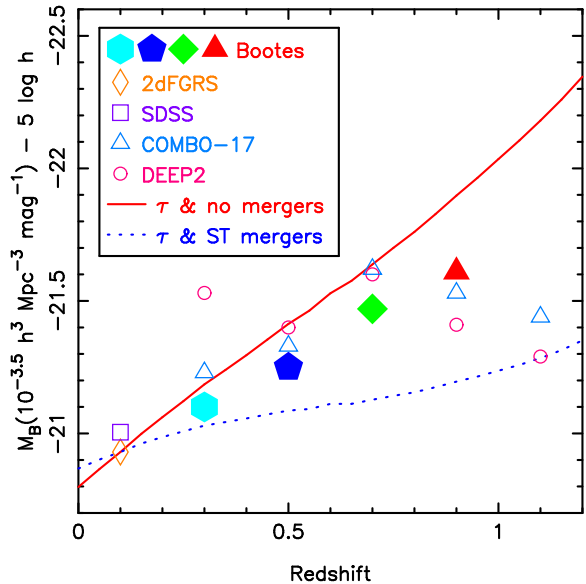}
\epsscale{1.00}
\caption{The luminosity evolution of $\simeq 4L^*$ red galaxies. We have parameterized
the evolution with $M_B(10^{-3.5})$, the absolute magnitude where the space density of 
red galaxies is $10^{-3.5}~h^3~{\rm Mpc}^{-3}~{\rm mag}^{-1}$. 
Random uncertainties for the \bootes~data are smaller than the size of the data points.
Simple models with and without stellar mass growth via galaxy mergers are plotted,
and these models are described in detail in \S\ref{sec:msec}.
Both models are normalized to the 2dFGRS and assume the stellar populations
of red galaxies fade by $1.24$ $B$-band magnitudes per unit redshift. 
Our simple merging model, based upon the \markcite{she99}{Sheth} \& {Tormen} (1999) mass function, underestimates
the space density of $4L^*$ red galaxies at $z>0.5$. While our model without mergers does not
fit the data, it is only offset from our data by $0.21$ magnitudes at $z=0.7$,
and we therefore conclude that $\simeq 80\%$ of the stellar mass contained within today's
$\simeq4L^*$ red galaxies was already in place at $z=0.7$.
\label{fig:m}}
\end{figure*}

We avoid these issues by measuring the evolving luminosity function at a 
fixed space density threshold. We parameterize the luminosity evolution 
with $M(10^{-3.5})$, the absolute magnitude where the space density of red 
galaxies is $10^{-3.5}~h^{-3}~{\rm Mpc}^{-3}~{\rm mag}^{-1}$. 
The most luminous red galaxies largely lie along the color magnitude relation,
so $M(10^{-3.5})$ is insensitive to details of red galaxy selection
criteria and it is thus easy to compare the results of various surveys.
Schechter functions provide a good fit to the red galaxy luminosity
function when the space density is above $10^{-4.5}~h^{-3}~{\rm Mpc}^{-3}~{\rm mag}^{-1}$
\markcite{mad02}(e.g., {Madgwick} {et~al.} 2002), so we use best-fit Schechter functions to derive
$M(10^{-3.5})$ from our data and the literature (when available).
Our estimates of $M(10^{-3.5})$ are provided in Table~\ref{table:mlf}, and plotted
in  Figure~\ref{fig:m} along with values derived from the literature.
We find $M_B(10^{-3.5})$ steadily increases with increasing redshift.
A straight line fit to the \bootes~ data alone has a reduced $\chi^2$ of $1.1$
and $M_B(10^{-3.5})$ brightens by $0.87\pm0.06$ $B$-band magnitudes from $z=0$ to 
$z=1$. While there are some discrepancies between the surveys at 
$z\sim 0.9$, which we discuss in \S\ref{sec:litcomp}, there is broad agreement 
between our results and the literature at $z<0.8$. If we fit to 
both the \bootes~ and 2dFGRS data, we find $M(10^{-3.5})$ brightens by 
$0.87\pm0.05$ $B$-band magnitudes between $z=0$ and $z=1$.

We use two simple models to characterize the evolution of $M_B(10^{-3.5})$. 
Both models are intended to be illustrative rather than precise descriptions 
of galaxy evolution.
Our simplest model has no galaxy mergers and assumes that the star formation 
history of all red galaxy progenitors is described by a \markcite{bru03}{Bruzual} \& {Charlot} (2003) 
$\tau=0.6~{\rm Gyr}$ model. This $\tau$ model has negligible star 
formation at $z<1$ and approximates the evolution of the  color-magnitude
and size-luminosity relations of $\sim 4L^*$ red galaxies (Figures~\ref{fig:uvz} and~\ref{fig:ap46}). 
As mergers of red galaxies do occur at $z<1$ \markcite{lau88,van05}(e.g., {Lauer} 1988; {van Dokkum} 2005), this 
model provides an upper bound for the fading of $M_B(10^{-3.5})$ from $z=1$ to $z=0$.

Our second model of $M_B(10^{-3.5})$ assumes galaxy mergers can be described by the \markcite{she99}{Sheth} \& {Tormen} (1999) 
halo mass function, which is similar to the formalism of \markcite{pre74}{Press} \& {Schechter} (1974).
We assume one galaxy per dark matter halo, a constant ratio of baryonic
matter to dark matter in each halo, and a $\tau=0.6~{\rm Gyr}$ model
star formation history. From the halo mass function we can determine the
space density of halos more massive than $m_{min}$, and we know the space
density of galaxies brighter than $M_B(10^{-3.5})$ is $\simeq 10^{-4} h^3 {\rm Mpc}^{-3}$. 
We therefore assume that the growth of baryons within $M_B(10^{-3.5})$ galaxies is 
proportional to the evolving $m_{min}$ value corresponding to a halo space density of 
$10^{-4} h^3 {\rm Mpc}^{-3}$. We find $m_{min}>10^{13} M_\odot$ at $z<1$, 
which is comparable to the masses of galaxy groups. 
This simple model is clearly flawed, as group and cluster halos contain multiple galaxies, and the 
infall of galaxies into these halos may not result in galaxy mergers.
As this model overestimates the rate of galaxy mergers, 
it should provide a lower bound for the evolution of $M_B(10^{-3.5})$.

Our $M_B(10^{-3.5})$ models are plotted in Figure~\ref{fig:m}, along with $M_B(10^{-3.5})$ 
values derived from this work and the literature. The model without mergers clearly provides 
the better approximation to the data. This model is offset from the data by only $0.21$ 
magnitudes at $z=0.7$ and $0.31$ magnitudes at $z=0.9$. If the stellar populations of
red galaxies fade in the same manner as the $\tau=0.6~{\rm Gyr}$ model, $\simeq 75\%$ 
of the stellar mass contained within today's $4L^*$ galaxies was already in place within 
these galaxies by $z=0.9$. If red galaxy stellar populations fade by less than 
$1.24$ $B$-band magnitudes from $z=1$ to $z=0$, an even higher percentage of the stellar 
mass of $4L^*$ red galaxies is already in place prior to $z=0.9$, and vice versa.
While we measure a higher space density of very luminous red galaxies at $z=0.9$
than the literature, there is broad agreement at lower redshifts. Even if the 
\bootes~measurements at $z=0.9$ are in error, the space density of $4L^*$ red galaxies from 
the literature is approximated by our model at $z\leq 0.7$. Roughly 80\% of the stellar mass 
within today's $4L^*$ red galaxies must already be in place by $z=0.7$.

We can see in Figure~\ref{fig:m} that the simple merging model, based upon the \markcite{she99}{Sheth} \& {Tormen} (1999) halo mass 
function for a $\Lambda{\rm CDM}$ cosmology, differs significantly from the 
observed evolution of  $M(10^{-3.5})$.
The simplifying assumptions used in this model may explain the offset from the data.
We assumed one galaxy per halo, while the most luminous red galaxies often reside 
within galaxy groups and clusters. An accurate model of these galaxies requires a  
description of how these galaxies occupy subhalos within groups and clusters, and 
how these subhalos merge over cosmic time. Such a model must also have 
negligible star formation at $z<1$, and clearly there is a physical process 
that is extremely effective at truncating star formation in $4L^*$ red galaxies.
Issues such as the halo occupation distribution function \markcite{ber02}(e.g., {Berlind} \& {Weinberg} 2002), gas 
heating \markcite{naa05}(e.g., {Naab} {et~al.} 2005), and AGN feedback \markcite{bow05,cro06,hop06}(e.g., {Bower} {et~al.} 2006; {Croton} {et~al.} 2006; {Hopkins} {et~al.} 2006) are 
clearly important for understanding galaxy evolution and need to be studied in more detail.

\section{COMPARISON TO OTHER RED GALAXY LUMINOSITY FUNCTIONS}
\label{sec:litcomp}

\begin{figure*}
\epsscale{0.90}
\plotone{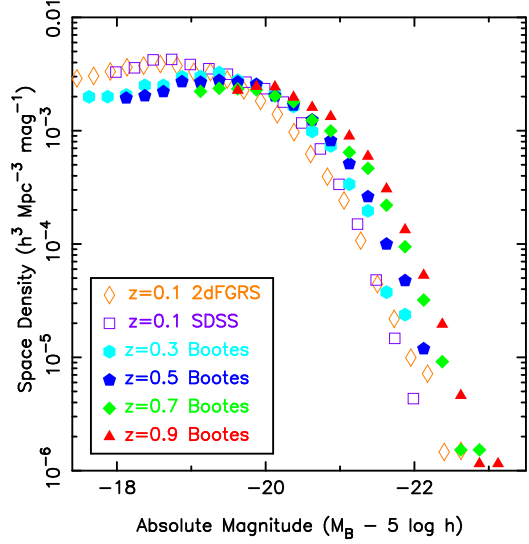}
\epsscale{1.00}
\caption{A comparison of our luminosity functions with those from the 
2dFGRS and SDSS \markcite{mad02,bla05}({Madgwick} {et~al.} 2002; {Blanton} 2005). To plot the 2dFGRS and SDSS luminosity 
functions, we have adopted $B_J-B=0.15$ and $g^{0.1}_{\rm AB}=B_{\rm Vega}$. 
There is a steady brightening of $>L^*$ red galaxies with increasing redshift.
\label{fig:lf_lowz}}
\end{figure*}

\begin{figure*}
\vspace{1cm}
\epsscale{0.90}
\plotone{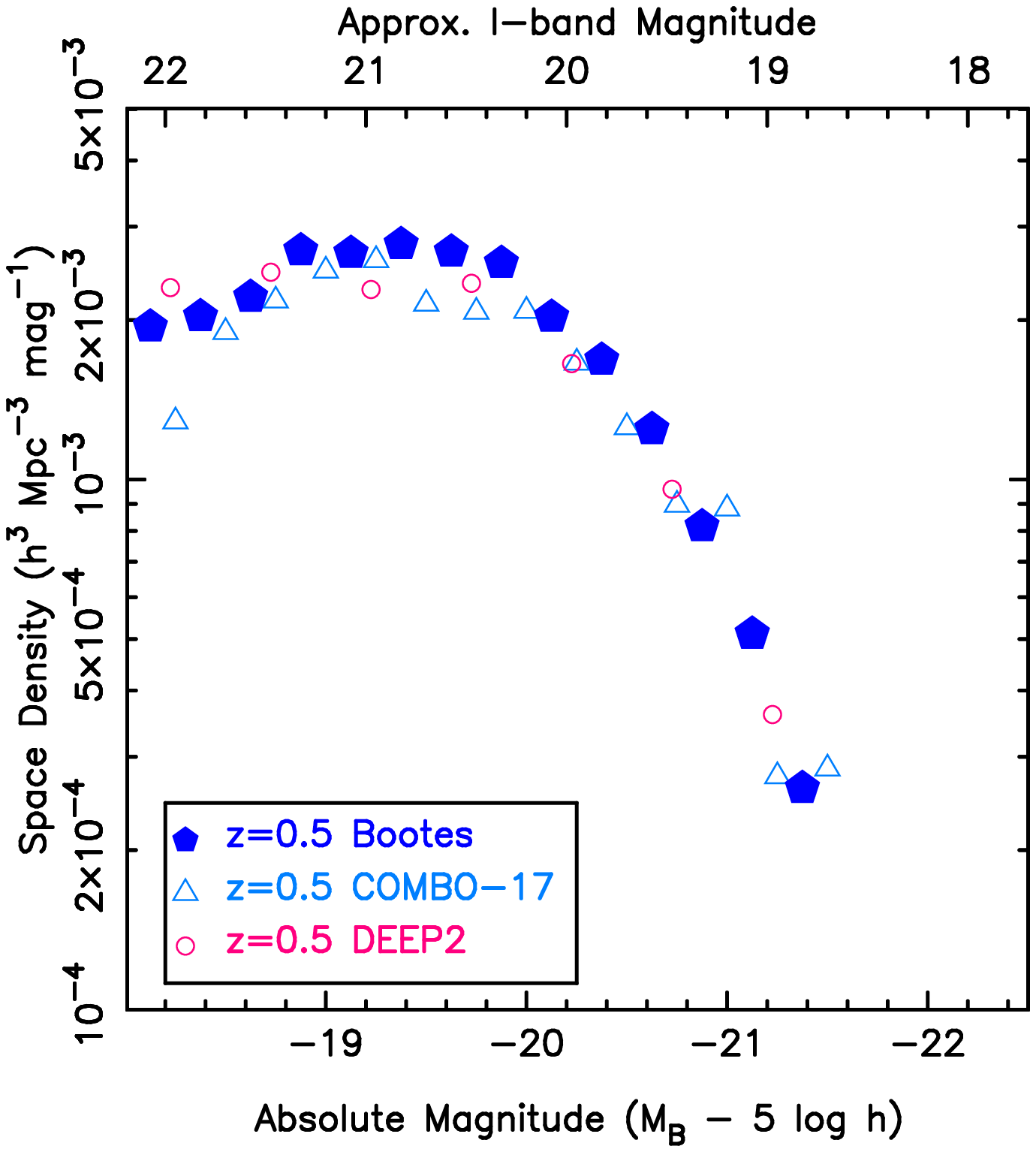}\plotone{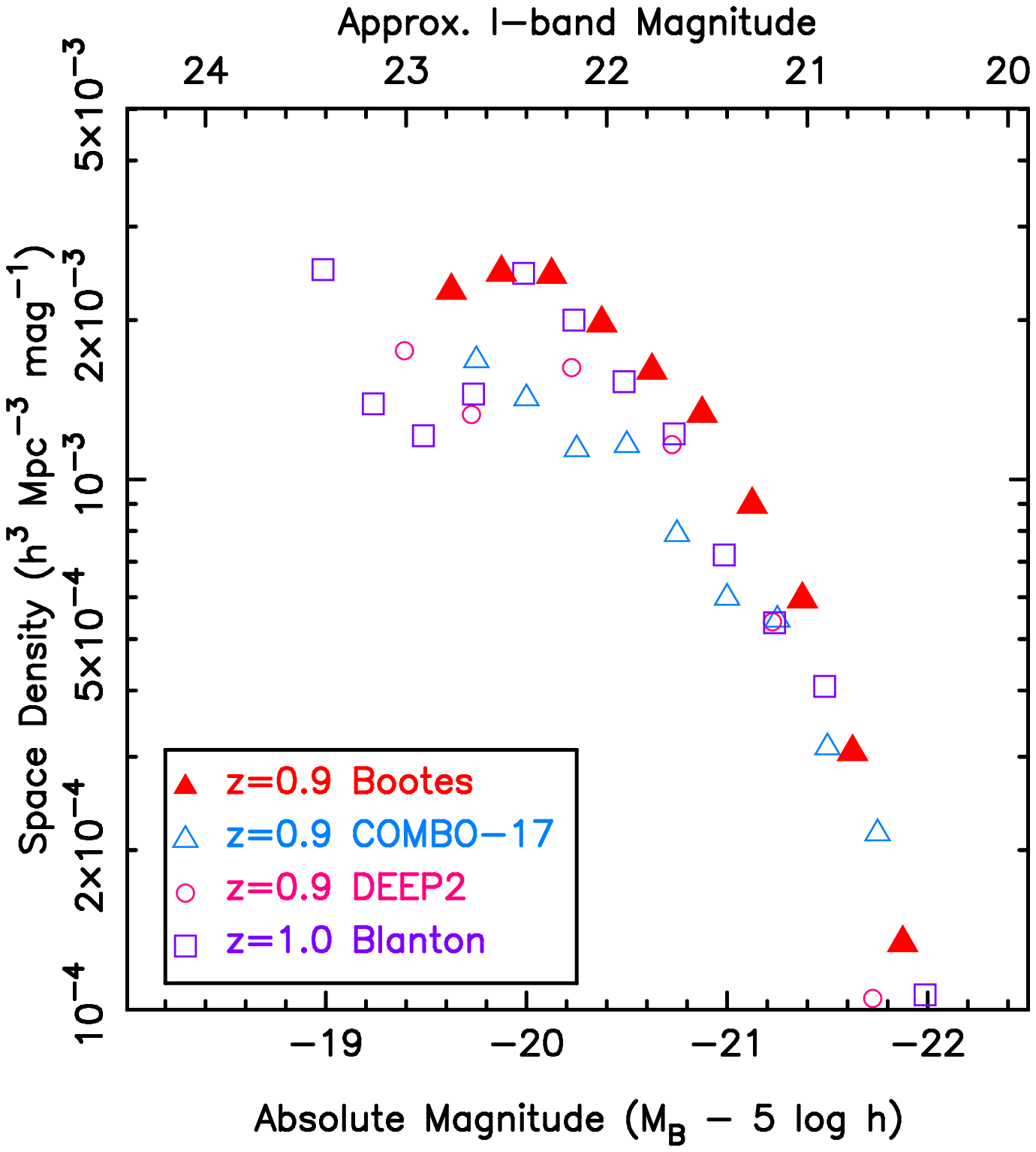}
\epsscale{1.00}
\caption{A comparison of our luminosity functions with those of 
COMBO-17 \markcite{fab05}({Faber} {et~al.} 2005), DEEP2 \markcite{wil05,fab05}({Willmer} {et~al.} 2005; {Faber} {et~al.} 2005), and the reanalysis of 
DEEP2 by \markcite{bla05}{Blanton} (2005). For clarity we don't plot the VVDS luminosity function of \markcite{zuc05}{Zucca} {et~al.} (2005), 
as their non-evolving red galaxy selection criterion differs greatly from the others displayed in this
 plot. The luminosity functions broadly agree at $z=0.5$, but at $z=0.9$ there
are significant disagreements between each of the surveys, particularly for $I>21$ red galaxies.
Potential causes for these discrepancies are discussed in \S\ref{sec:litcomp}.
\label{fig:lf_highz}}
\end{figure*}

There have been a variety of conclusions about red galaxy evolution 
over the past decade \markcite{fab05}(see  {Faber} {et~al.} 2005), and some of this can be 
explained by the different behavior of $L^*$ and $4L^*$ red galaxies 
(\S\ref{sec:jbsec} and \S\ref{sec:msec} respectively).
\markcite{bun05}{Bundy} {et~al.} (2005) also find that the evolution of red galaxies is 
a function of stellar mass, and this has been confirmed by recent
analyzes of COMBO-17 stellar mass and luminosity functions \markcite{bor06,cim06}({Borch} {et~al.} 2006; {Cimatti}, {Daddi}, \& {Renzini} 2006). 
The luminosity density of red galaxies is heavily weighted towards 
$L^*$ galaxies and evolves slowly at $z<1$. After accounting for fading 
stellar populations, we and much of the recent literature \markcite{bel04,fab05}(e.g., {Bell} {et~al.} 2004; {Faber} {et~al.} 2005)
conclude that the stellar mass contained within the ensemble of red galaxies has steadily 
increased since $z=1$. In contrast, the evolution of $4L^*$ red galaxies 
differs only slightly from pure luminosity evolution. However, unlike some recent
studies \markcite{bun05,bor06,cim06}(e.g., {Bundy} {et~al.} 2005; {Borch} {et~al.} 2006; {Cimatti} {et~al.} 2006), we do see evidence for the ongoing assembly
of the most massive red galaxies, albeit at a rate that produces little
growth of $4L^*$ red galaxy stellar masses at $z<1$.
As red galaxy evolution is a function of luminosity, it is not surprising that 
various authors using different techniques 
have reached a variety of conclusions, even when using the same galaxy surveys 
\markcite{bel04,wil05,fab05,bla05,bun05,bor06,cim06}(e.g., {Bell} {et~al.} 2004; {Willmer} {et~al.} 2005; {Faber} {et~al.} 2005; {Blanton} 2005; {Bundy} {et~al.} 2005; {Borch} {et~al.} 2006; {Cimatti} {et~al.} 2006).

Figures~\ref{fig:lf_lowz} and~\ref{fig:lf_highz} show our evolving luminosity 
functions are broadly consistent with the 2dFGRS, SDSS, COMBO-17 and DEEP2 at 
$z\lesssim 0.8$. There is a steady increase of the bright end of the luminosity 
function with redshift. If we combine our results with those of the 2dFGRS and SDSS, 
there is a gradual decline of  $\phi^*$ with increasing redshift.
However, we show in Figure~\ref{fig:lf_highz} that we measure a higher space 
density of $z=0.9$ red galaxies than DEEP2 or COMBO-17. Since these studies 
have had a significant impact upon the field, it is important to understand why 
these differences occur.

\begin{figure}[hbt]
\epsscale{0.85}
\plotone{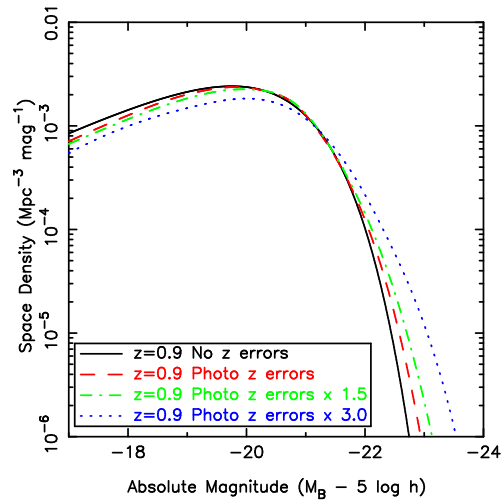}
\epsscale{1.00}
\caption{The effect of photometric redshift errors on the measured $z=0.9$ luminosity function.
We use the photometric redshift uncertainties listed in Table~\ref{table:photoz}, except when there
are ten or less spectroscopic redshifts, in which case we extrapolate from brighter magnitudes
by assuming the uncertainties increase by 50\% per unit magnitude.
The effect of photometric redshift errors upon the luminosity function is only significant
for $M_B-5{\rm log} h<-22.5$ galaxies, where the space density is overestimated.
In Figure~\ref{fig:lfz} the excess of very luminous galaxies compared to a Schechter 
function may be caused by photometric redshift errors. 
Even if our photometric redshift uncertainties are 50\% larger those listed in Table~\ref{table:photoz}, the impact on our
results and conclusions is small as the measured luminosity function of $-21.8<M_B-5{\rm log} h<-19.5$ red galaxies
remains largely unchanged.
\label{fig:lfscat}}
\end{figure}

An obvious difference between our work and some of the recent literature
is our use of photometric redshifts. Photometric redshift errors 
can scatter numerous low luminosity objects into high luminosity bins,
and thus alter the shape of the luminosity function  \markcite{bro01,che03}(e.g., {Brown}, {Webster}, \& {Boyle} 2001; {Chen} {et~al.} 2003). 
To account for this we convolved our evolving luminosity function by the 
expected luminosity and volume errors resulting from our photometric 
redshifts. We use the photometric redshift errors listed in 
Table~\ref{table:photoz}, except at faint magnitudes with ten or less
spectroscopic redshifts, where we extrapolate from brighter magnitudes by 
assuming the errors increase by 50\% per unit magnitude.
In Figure~\ref{fig:lfscat} the largest error is a
$\simeq 0.2$ magnitude offset for rare $10L^*$ galaxies at $z=0.9$. 
In Figure~\ref{fig:lfz}, there are slightly more  $10L^*$ galaxies 
than expected from our best-fit $z=0.9$ Schechter function, but these 
are a tiny fraction of our sample and do not affect our results, 
which rely upon less luminous galaxies.

Compared to the other luminosity functions in Table~\ref{table:mlf}, 
the VIRMOS-VLT Deep Survey \markcite{zuc05}(VVDS; {Zucca} {et~al.} 2005) find a lower space density 
of high redshift and low luminosity red galaxies.
As discussed by  \markcite{zuc05}{Zucca} {et~al.} (2005), this is not unexpected as the VVDS 
selection criteria do not model the evolution and tilt of the color-magnitude relation. 
\markcite{wol03}{Wolf} {et~al.} (2003) also observed a rapid decline of $\phi^*$ with redshift for a sample 
of red galaxies selected from COMBO-17 with a non-evolving criterion.
Similar selection effects may occur in other surveys by accident at $z>1$, where the 
evolving color-magnitude relation is relatively difficult to measure and red 
galaxy selection criteria could unintentionally intercept the color-magnitude relation.

Both the VVDS and COMBO-17 use SExtractor MAG\_AUTO to determine total magnitudes, and 
accuracy of the SDSS DR1 and 2dFGRS photometry has been verified with deep CCD photometry 
making use of MAG\_AUTO \markcite{cro04}({Cross} {et~al.} 2004). MAG\_AUTO is known to underestimate the luminosities
of galaxies with \markcite{dev48}{de Vaucouleurs} (1948) profiles by $0.1$ magnitudes when the half-light radius is several
times larger than the stellar full width at half maximum \markcite{cro04}({Cross} {et~al.} 2004). It is therefore plausible 
that the brightest red galaxies in the 2dFGRS have had their luminosities underestimated by 
$\sim 10\%$. As illustrated in Figure~\ref{fig:sys}, MAG\_AUTO can also have systematics of several 
tenths of a magnitude at $I>21$, even when the catalogs are 85\% complete. We therefore speculate 
that the VVDS and COMBO-17 may have systematic errors on the order of $0.2$ magnitudes at the 
faint end of their luminosity functions.

While our $z\simeq 0.9$ luminosity functions and those derived from 
DEEP2 \markcite{wil05,fab05,bla05}({Willmer} {et~al.} 2005; {Faber} {et~al.} 2005; {Blanton} 2005) are broadly similar in shape and normalization, 
there is a systematic offset of $\simeq 0.2$ magnitudes. This offset is present even 
at high luminosities, which are largely insensitive to the selection effects
discussed elsewhere in this paper. As discussed in \S\ref{sec:rest}, we correct 
for flux outside of our aperture using a model derived from the size-luminosity 
relation. Such corrections are typically not applied to samples in the literature.

\begin{figure}[hbt]
\epsscale{0.85}
\plotone{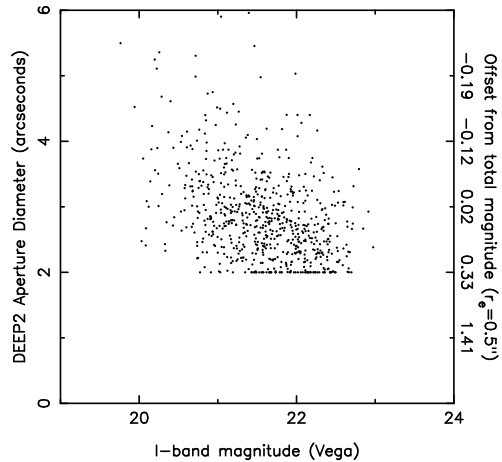}
\epsscale{1.00}
\caption{The DEEP2 aperture diameter for $R-I>1.2$, $0.8<z<1.0$ galaxies from 
the first DEEP2 data release. The apparent color cut effectively selects galaxies
with red rest-frame colors in this redshift range. On the right we list the expected
magnitude offsets for galaxies with a $0.5^{\prime\prime}$ half-light radius observed in 
$0.8^{\prime\prime}$ seeing and measured with the DEEP2 aperture. 
As the DEEP2 photometry is zero-pointed with stars measured in a  
$2^{\prime\prime}$ aperture, larger apertures overestimate 
the luminosities of point sources and very large apertures can overestimate 
galaxy luminosities. 
However, we find that DEEP2 photometry typically underestimates the total
luminosities of $z \simeq 0.9$ red galaxies.
\label{fig:deepap}}
\end{figure}

\begin{figure*}[hbt]
\vspace{1cm}
\epsscale{0.75}
\plotone{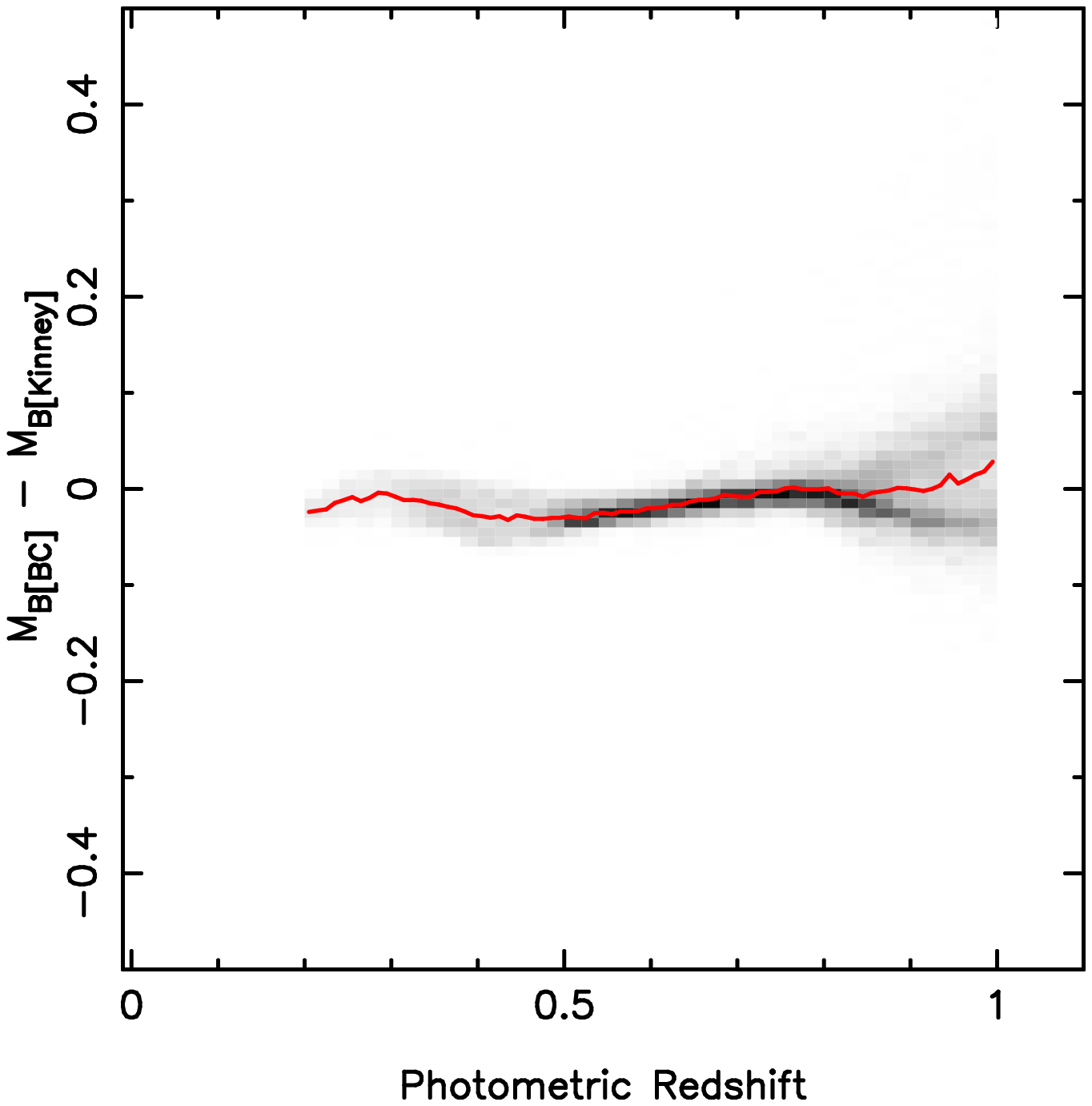}\plotone{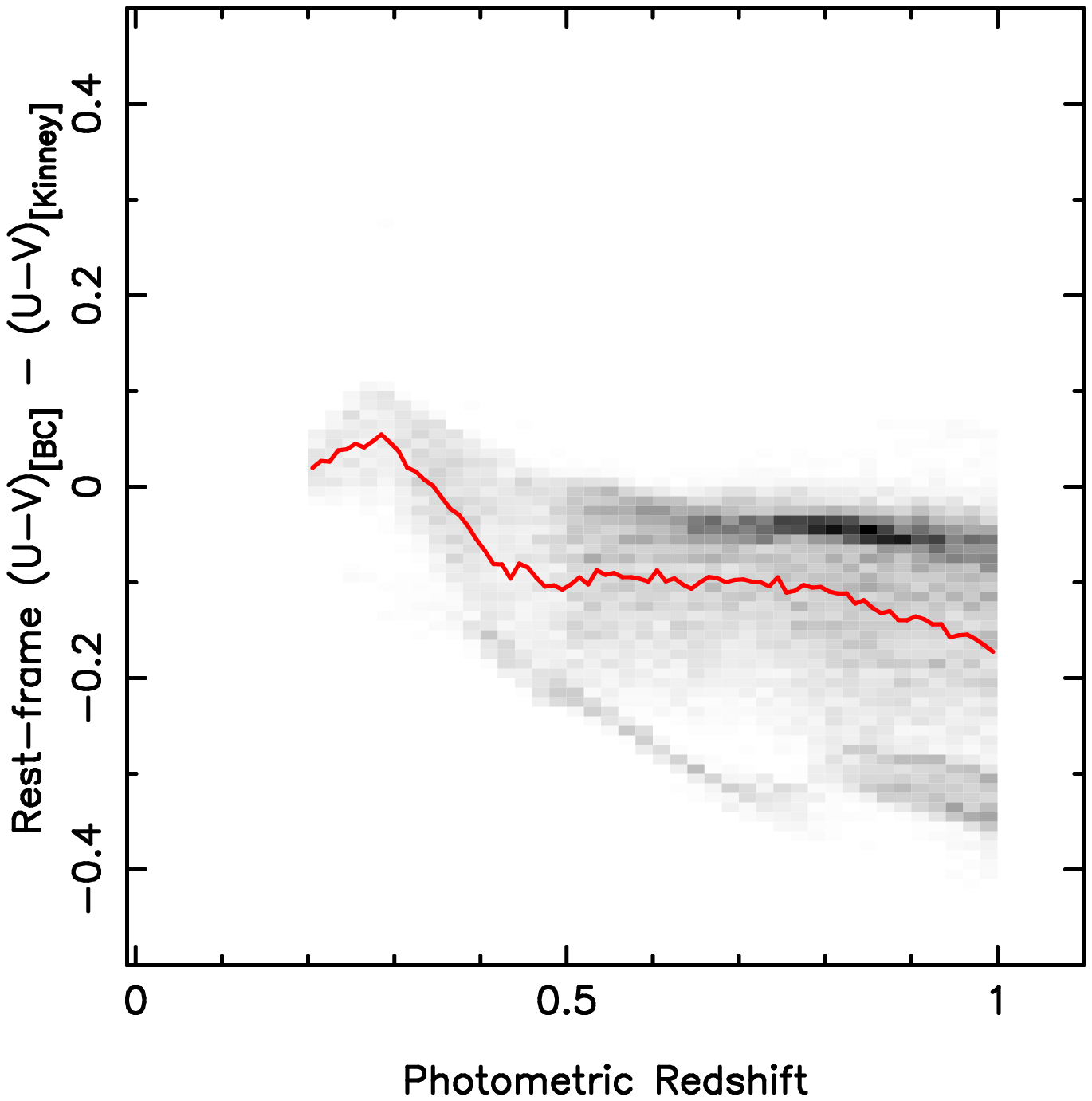}
\epsscale{1.00}
\caption{The difference between absolute magnitudes and rest-frame colors determined with
\markcite{bru03}{Bruzual} \& {Charlot} (2003) models and \markcite{kin96}{Kinney} {et~al.} (1996) templates. For this plot we have only included 
$M_B - 5 {\rm log}h <-19.5$ red galaxies.
To produce a continuous model of the galaxy locus, we interpolated between the \markcite{kin96}{Kinney} {et~al.} (1996) SEDs of 
elliptical, S0, Sa, Sb, and Sc galaxies.
The red solid line shows the mean offset as a function of redshift.
While the $B$-band absolute magnitudes are relatively insensitive to the choice of 
SED models or templates, the rest-frame $U-V$ colors show systematics on the 
order of 0.2 magnitudes. If we have overestimated the evolution of $U-V$ with redshift,
our principal conclusions remain unchanged. 
\label{fig:kin}}
\end{figure*}

We have estimated the fraction of galaxy flux that falls beyond the photometric
aperture used by the DEEP2 survey. DEEP2 photometry uses the larger of a 
$2^{\prime\prime}$ diameter aperture or a $6\sigma$ aperture, where $\sigma$ is 
defined by a Gaussian fit to the object profile \markcite{coi04}({Coil} {et~al.} 2004). DEEP2 photometry 
is zero-pointed with SDSS photometry of stars, where the SDSS uses PSF magnitudes while DEEP2 
uses apertures of $\simeq 2^{\prime\prime}$ diameter. Because of this, 
a point source measured with a $>2^{\prime\prime}$ diameter aperture
will have its flux systematically overestimated. Similarly, the photometry
of galaxies will have systematic errors which are a function of half-light radius and 
aperture size. In Figure~\ref{fig:deepap}, we plot the DEEP2 aperture diameter as a
function of apparent magnitude for $0.8<z<1.0$ red galaxies selected
from the first DEEP2 data release. At these redshifts, an $I\sim 21.5$
galaxy has a half-light radius of $\simeq 0.5^{\prime\prime}$ ($\simeq 3 h^{-1}{\rm kpc}$),
and we list the systematic error for such a galaxy
measured with the DEEP2 aperture on the right of  Figure~\ref{fig:deepap}.
Red galaxies at $z\sim 0.9$ may have their luminosities systematically underestimated 
by $\simeq 0.15$ magnitudes by DEEP2, and we therefore conclude that 
different measures of total magnitudes contribute to the difference
between our work and the prior literature.

Different surveys use different SED models and templates to derive rest-frame 
galaxy properties from observed photometry.  
Figure~\ref{fig:kin} compares rest-frame magnitudes and colors derived with 
the \markcite{bru03}{Bruzual} \& {Charlot} (2003) $\tau$ models and \markcite{kin96}{Kinney} {et~al.} (1996) templates for our red galaxy sample. 
The absolute magnitudes exhibit small differences, as the observed $I$-band
is very close to the rest-frame $B$-band at $z=0.9$. 
In contrast, the rest-frame $U-V$ colors show systematics as large as 
0.2 magnitudes. This will have minimal impact on the bright end of the luminosity 
function, as the most luminous red
galaxies lie along the color-magnitude relation, which can be identified empirically.
The greatest impact from SED errors will be at lower luminosities, as
SED errors will shift red galaxy selection criterion relative to the 
the color-magnitude relation. A $0.1$ magnitude shift in our $U-V$ selection criterion
results in our $\phi^*$ and $j_B$ values changing by $\sim 25\%$.

If the \markcite{kin96}{Kinney} {et~al.} (1996) templates are better models of $z\sim 0.9$ red galaxy SEDs
than the \markcite{bru03}{Bruzual} \& {Charlot} (2003) $\tau$ models, our principal conclusions remain unchanged.
The $U-V$ color-magnitude relation still evolves, but the mean color and luminosity
of red galaxy stellar populations evolves less rapidly than predicted by the $\tau=0.6~{\rm Gyr}$ 
model. Even if our $z=0.9$ $j_B$ measurement has a $25\%$ error, the stellar 
mass contained within the red galaxy population must increase with decreasing redshift.
If we reduce the luminosity evolution of red galaxy stellar populations, 
a revised $M_B(10^{-3.5})$ model without mergers will provide a better 
fit to the observations in Figure~\ref{fig:m} than our current model.

\section{SUMMARY}
\label{sec:sum}

We have measured the evolution and assembly of red galaxies over 
the last $8~{\rm Gyr}$, using the evolving  $B$-band luminosity 
function of $0.2<z<1.0$ red galaxies. 
Our sample of 39599 $0.2<z<1.0$ red galaxies, selected from 
$6.96~{\rm deg}^2$ of imaging, is an order of magnitude larger in 
size and volume than comparable $z\sim 0.9$ red galaxy samples. Accurate 
photometric redshifts were determined using the empirical ANNz photometric 
redshift code and photometry from both the NOAO Deep Wide-Field and {\it Spitzer} 
IRAC shallow surveys. The accuracy of these redshifts has been verified with 
spectroscopy, and is comparable to the best broad-band photometric redshifts 
in the literature.

We find the color-magnitude and size-luminosity relations of 
red galaxies evolve with redshift, and this evolution can be 
approximated by a \markcite{bru03}{Bruzual} \& {Charlot} (2003) $\tau=0.6~{\rm Gyr}$ stellar 
population synthesis model. The size-luminosity relation predicts 
a significant fraction of the galaxy flux is at radii of 
several arcseconds, even at $z\sim 0.9$. Luminosity functions 
which do not account for this underestimate the space density 
of red galaxies at high redshift, and overestimate the assembly rate 
of red galaxies at $z<1$. 

We find the luminosity density, $j_B$, of red galaxies increases 
by $36\pm13\%$ from $z=0$ to $z=1$. In contrast, the $B$-band 
luminosity of a $\tau=0.6~{\rm Gyr}$ stellar population model increases by 
213\% over the same redshift range. If red galaxy stellar populations 
fade by $1.24$ magnitudes from $z=1$ to $z=0$, the stellar mass contained 
within red galaxies has approximately doubled over the past $8~{\rm Gyr}$.
Blue galaxies are being transformed into $\lesssim L^*$ red galaxies at $z<1$, after 
a steady decline or rapid truncation of their star formation.

The evolution of the most luminous red galaxies at $z<1$ cannot be dominated by 
fading blue galaxies, as blue galaxies with large stellar masses are exceptionally 
rare at $z<1$. We have measured the evolution of $4L^*$ red galaxies
using $M_B(10^{-3.5})$, the absolute magnitude corresponding to a fixed space
density of $10^{-3.5}~h^3~{\rm Mpc^{-3}}~{\rm mag^{-1}}$.
Our measurements of $M_B(10^{-3.5})$, along with those we derived from the literature, 
show that the stellar masses of the most luminous red galaxies evolve slowly at $z<1$.
A \markcite{bru03}{Bruzual} \& {Charlot} (2003) $\tau=0.6~{\rm Gyr}$ model, with little stellar mass evolution and 
normalized to the 2dFGRS, only overestimates $M_B(10^{-3.5})$ by $\simeq 0.21$ magnitudes at $z=0.7$. 
We therefore conclude that $\simeq 80\%$ of the stellar mass contained within today's $M_B-5{\rm log}h=-21$
red galaxies was already in place within these galaxies at $z=0.7$. While red galaxy mergers have been
reported in the prior literature, such mergers do not rapidly increase the stellar masses of $4L^*$ red galaxies 
between $z=1$ and the present day.

\acknowledgments

We thank our colleagues on the NDWFS, IRAC Shallow Survey, and AGES teams, 
in particular R.~J. Cool, D.~J. Eisenstein, G.~G. Fazio, C.~S. Kochanek, S.~S. Murray, and G.~P. Tiede.
This paper would not have been possible without the efforts of the  KPNO, {\it Spitzer}, MMT, 
W.~M.~Keck and {\it Gemini} support staff.
We are grateful to the IRAF team for the majority of the packages used to process the NDWFS images. 
We thank Alyson Ford, Lissa Miller, and Jennifer Claver, for reducing much of the NDWFS data used for this paper. 
H. Spinrad, S. Dawson, D. Stern, J.~E. Rhoads, S Malhotra, B.~T. Soifer, C. Bian, S.~G. Djorgovski, S.~A. Stanford, 
S.~Croft, W.~van~Breugel and the AGES collaboration generously shared their spectroscopic redshifts with us prior to publication.
This work is based in part on observations made with the {\it Spitzer} Space Telescope, which is operated 
by the Jet Propulsion Laboratory, California Institute of Technology 
under a contract with NASA. 
This research was supported by the National Optical Astronomy Observatory which is
operated by the Association of Universities for Research in Astronomy (AURA), Inc.
under a cooperative agreement with the National Science Foundation.
Most of the spectroscopic redshifts discussed in this paper were obtained at the MMT Observatory, a 
joint facility of the Smithsonian Institution and the University of Arizona.
While writing this paper we had many productive discussions with other astronomers working upon galaxy
assembly and evolution, including E.~F. Bell, M.~R.~Blanton, A.~L.~Coil, S.~M.~Faber, 
J.~.E.~Gunn, T.~R.~Lauer, J.~A. Newman, J.~P. Ostriker, C.~N.~A. Willmer, C.~Wolf, and E.~Zucca.


\bibliography{}

\begin{deluxetable}{cccccccccccccccc}
\tablecolumns{13}
\tabletypesize{\scriptsize}
\tablecaption{Measured Photometric Redshift Uncertainties For Red Galaxies\label{table:photoz}}
\tablehead{ 
\colhead{}           & 
\multicolumn{3}{c}{$0.2<z<0.4$}  &
\multicolumn{3}{c}{$0.4<z<0.6$}  &
\multicolumn{3}{c}{$0.6<z<0.8$}  &
\multicolumn{3}{c}{$0.8<z<1.0$} \\
\colhead{$4^{\prime\prime}$ Apparent} & 
\multicolumn{3}{c}{$M_B-5{\rm log}h<-17.5$} & 
\multicolumn{3}{c}{$M_B-5{\rm log}h<-18.0$} & 
\multicolumn{3}{c}{$M_B-5{\rm log}h<-19.0$} & 
\multicolumn{3}{c}{$M_B-5{\rm log}h<-19.5$} \\
\colhead{Magnitude} &
68.7\% & 90\% & $N$ & 
68.7\% & 90\% & $N$ & 
68.7\% & 90\% & $N$ & 
68.7\% & 90\% & $N$ 
}
\startdata
 $ 17.0<I<18.0 $   &  0.033 & 0.060 & $( 170)$  &  --    & --    & --        &  --    & --    & --        &  --    & --    & --        \\ 
 $ 18.0<I<19.0 $   &  0.032 & 0.063 & $(1110)$  &  0.026 & 0.051 & $( 132)$  &  0.025 & 0.025 & $(   1)$  &  --    & --    & --        \\ 
 $ 19.0<I<20.0 $   &  0.039 & 0.074 & $(1007)$  &  0.032 & 0.058 & $( 949)$  &  0.029 & 0.053 & $( 192)$  &  0.027 & 0.179 & $(  10)$  \\ 
 $ 20.0<I<21.0 $   &  0.049 & 0.088 & $( 166)$  &  0.036 & 0.066 & $( 340)$  &  0.040 & 0.072 & $( 240)$  &  0.058 & 0.096 & $(  50)$  \\ 
 $ 21.0<I<22.0 $   &  0.042 & 0.363 & $(   4)$  &  0.229 & 3.066 & $(   5)$  &  0.103 & 0.131 & $(  14)$  &  0.045 & 0.090 & $(  28)$  \\ 
 $ 22.0<I<23.5 $   &  --    & --    & --        &  --    & --    & --        &  0.162 & 0.271 & $(   5)$  &  0.127 & 0.254 & $(   7)$  \\ 
\enddata
\end{deluxetable}

\begin{deluxetable}{ll}
\tablecolumns{2}
\tabletypesize{\scriptsize}
\tablecaption{Color cuts used to exclude stars, quasars and $z\gg 1$ galaxies from the sample.\label{table:colcut}}
\tablehead{ 
\colhead{Magnitude and/or color range} &
\colhead{Cut}    
}
\startdata
$B_W\leq 26.5$                                 & $B_W-R<1.00$                          \\
$B_W\leq 26.5$                                 & $B_W-R>4.20$                          \\
$B_W>26.5$                                     & $26.5-R>4.20$                         \\
$B_W\leq 26.5$                                 & $B_W-R>2.55+2.75\times(R-I-0.55)$     \\
$B_W>26.5$                                     & $26.5-R>2.55+2.75\times(R-I-0.55)$    \\
$I<23.5$                                       & $R-I<0.55$                            \\
$I<23.5$                                       & $R-I>1.70$                            \\
$B_W\leq 26.5$                                 & $B_W-R<2.0-(R-I)$                     \\
$B_W\leq 26.5$                                 & $B_W-R<5.0\times(R-I-1.2)$            \\
$[3.6]\leq 19.5$                               & $I-[3.6]<2.2$                         \\
$[3.6]>19.5$                                   & $I-19.5<2.2$                          \\
$[3.6]\leq 19.5$ and $0.75\leq R-I<1.05$       & $R-I>0.75+0.375\times(I-[3.6]-2.2)$   \\
$[3.6]>19.5$     and $0.75\leq R-I<1.05$       & $R-I>0.75+0.375\times(I-19.5-2.2)$    \\
$[3.6]\leq 19.5$ and $1.05\leq R-I<1.70$       & $R-I>1.05+0.60\times(I-[3.6]-3.0)$    \\
$[3.6]>19.5$     and $1.05\leq R-I<1.70$       & $R-I>1.05+0.60\times(I-19.5-3.0)$     \\
$[3.6]\leq 19.5$ and $[3.6]\leq 18.5$          & $[3.6]-[4.5]>0.6$                     \\
$[3.6]\leq 19.5$ and $[3.6]\leq 18.5$          & $[3.6]-[4.5]<-0.6$                    \\
\enddata
\end{deluxetable}

\begin{deluxetable}{cccccccccccccccc}
\tablecolumns{5}
\tabletypesize{\scriptsize}
\tablecaption{Red Galaxy Number Counts\label{table:counts}}
\tablehead{ 
\colhead{$4^{\prime\prime}$ Apparent}           & 
\multicolumn{1}{c}{$0.2<z<0.4$}  &
\multicolumn{1}{c}{$0.4<z<0.6$}  &
\multicolumn{1}{c}{$0.6<z<0.8$}  &
\multicolumn{1}{c}{$0.8<z<1.0$} \\
\colhead{Magnitude} & 
\multicolumn{1}{c}{$M_B-5{\rm log}h<-17.5$} & 
\multicolumn{1}{c}{$M_B-5{\rm log}h<-18.0$} & 
\multicolumn{1}{c}{$M_B-5{\rm log}h<-19.0$} & 
\multicolumn{1}{c}{$M_B-5{\rm log}h<-19.5$}
}
\startdata
 $ 17.0<I<18.0 $    &   189   & -----   & -----   & ----- \\ 
 $ 18.0<I<19.0 $    &  1272   &   141   &     1   & ----- \\ 
 $ 19.0<I<20.0 $    &  2365   &  2131   &   400   &    14 \\ 
 $ 20.0<I<21.0 $    &  1719   &  4591   &  3320   &   977 \\ 
 $ 21.0<I<22.0 $    &   438   &  3512   &  5738   &  5466 \\ 
 $ 22.0<I<23.5 $    & -----   &   608   &  1359   &  5358 \\ 
\enddata
\end{deluxetable}


\begin{deluxetable}{ccccccccc}
\tablecolumns{9}
\tabletypesize{\scriptsize}
\tablecaption{Comparison of subsample and clustering uncertainties for $\phi^*$ and $j_B$\label{table:clu}}
\tablehead{
\colhead{$z$ range}                                                                                  &
\colhead{$\omega(1^\prime)$}                                                                         &
\colhead{$\gamma$}                                                                                   &
\multicolumn{3}{c}{$\phi^* \times 10^2 (h^3 {\rm Mpc^{-3}~mag^{-1}})$}                                 &  
\multicolumn{3}{c}{$j_B (10^{7} h~L_{\odot}~{\rm Mpc^{-3}})$} \\
\colhead{} &
\colhead{} &
\colhead{} &
\colhead{Best fit} &
\colhead{Subsample} &
\colhead{Clustering} &
\colhead{Best fit} &
\colhead{Subsample} &
\colhead{Clustering} \\
\colhead{} &
\colhead{} &
\colhead{} &
\colhead{Value} &
\colhead{$1\sigma$} &
\colhead{$1\sigma$} &
\colhead{Value}     &
\colhead{$1\sigma$} &
\colhead{$1\sigma$} 
}
\startdata
$0.20<z<0.40$  & $ 0.74 \pm 0.05 $ &  $ 1.94 \pm 0.05 $  & $ 8.45 $ & $ \pm 0.38  $ & $ \pm 1.07 $   & $  8.1 $ & $ \pm  0.3  $ & $ \pm  1.0 $  \\ 
$0.40<z<0.60$  & $ 0.51 \pm 0.03 $ &  $ 2.02 \pm 0.07 $  & $ 7.61 $ & $ \pm 0.34  $ & $ \pm 0.68 $   & $  8.5 $ & $ \pm  0.4  $ & $ \pm  0.8 $  \\ 
$0.60<z<0.80$  & $ 0.28 \pm 0.02 $ &  $ 2.06 \pm 0.06 $  & $ 5.71 $ & $ \pm 0.33  $ & $ \pm 0.35 $   & $  9.2 $ & $ \pm  0.4  $ & $ \pm  0.6 $  \\ 
$0.80<z<1.00$  & $ 0.24 \pm 0.02 $ &  $ 1.93 \pm 0.05 $  & $ 6.35 $ & $ \pm 0.21  $ & $ \pm 0.47 $   & $ 10.7 $ & $ \pm  0.3  $ & $ \pm  0.8 $  \\ 
\enddata
\end{deluxetable}

\begin{deluxetable}{ccccc}
\tablecolumns{5}
\tabletypesize{\scriptsize}
\tablecaption{Red Galaxy $1/V_{max}$ Luminosity Function with subsample uncertainties\label{table:vmax}}
\tablehead{ 
\colhead{Absolute}            & 
\multicolumn{4}{c}{Luminosity Function ($h^3~{\rm Mpc^{-3}~mag^{-1}}$)} 
\\  
  \colhead{Magnitude}           & 
  \colhead{$0.2<z<0.4$}         &
  \colhead{$0.4<z<0.6$}         &
  \colhead{$0.6<z<0.8$}         &
  \colhead{$0.8<z<1.0$}         
}
\startdata
$ -17.75 < M_B-5{\rm log}h < -17.50 $ & $ 1.99 \pm 0.10 \times 10^{-3} $ & -                                & -                                & -                                \\ 
$ -18.00 < M_B-5{\rm log}h < -17.75 $ & $ 1.99 \pm 0.13 \times 10^{-3} $ & -                                & -                                & -                                \\ 
$ -18.25 < M_B-5{\rm log}h < -18.00 $ & $ 2.07 \pm 0.10 \times 10^{-3} $ & $ 1.94 \pm 0.12 \times 10^{-3} $ & -                                & -                                \\ 
$ -18.50 < M_B-5{\rm log}h < -18.25 $ & $ 2.51 \pm 0.18 \times 10^{-3} $ & $ 2.03 \pm 0.13 \times 10^{-3} $ & -                                & -                                \\ 
$ -18.75 < M_B-5{\rm log}h < -18.50 $ & $ 2.51 \pm 0.13 \times 10^{-3} $ & $ 2.21 \pm 0.11 \times 10^{-3} $ & -                                & -                                \\ 
$ -19.00 < M_B-5{\rm log}h < -18.75 $ & $ 2.98 \pm 0.17 \times 10^{-3} $ & $ 2.71 \pm 0.16 \times 10^{-3} $ & -                                & -                                \\ 
$ -19.25 < M_B-5{\rm log}h < -19.00 $ & $ 3.00 \pm 0.17 \times 10^{-3} $ & $ 2.68 \pm 0.13 \times 10^{-3} $ & $ 2.22 \pm 0.11 \times 10^{-3} $ & -                                \\ 
$ -19.50 < M_B-5{\rm log}h < -19.25 $ & $ 3.25 \pm 0.18 \times 10^{-3} $ & $ 2.78 \pm 0.11 \times 10^{-3} $ & $ 2.36 \pm 0.11 \times 10^{-3} $ & -                                \\ 
$ -19.75 < M_B-5{\rm log}h < -19.50 $ & $ 2.78 \pm 0.14 \times 10^{-3} $ & $ 2.69 \pm 0.10 \times 10^{-3} $ & $ 2.34 \pm 0.11 \times 10^{-3} $ & $ 2.27 \pm 0.08 \times 10^{-3} $ \\ 
$ -20.00 < M_B-5{\rm log}h < -19.75 $ & $ 2.55 \pm 0.14 \times 10^{-3} $ & $ 2.56 \pm 0.18 \times 10^{-3} $ & $ 2.29 \pm 0.10 \times 10^{-3} $ & $ 2.46 \pm 0.10 \times 10^{-3} $ \\ 
$ -20.25 < M_B-5{\rm log}h < -20.00 $ & $ 2.17 \pm 0.14 \times 10^{-3} $ & $ 2.03 \pm 0.12 \times 10^{-3} $ & $ 2.02 \pm 0.14 \times 10^{-3} $ & $ 2.44 \pm 0.07 \times 10^{-3} $ \\ 
$ -20.50 < M_B-5{\rm log}h < -20.25 $ & $ 1.64 \pm 0.08 \times 10^{-3} $ & $ 1.68 \pm 0.07 \times 10^{-3} $ & $ 1.80 \pm 0.06 \times 10^{-3} $ & $ 1.97 \pm 0.07 \times 10^{-3} $ \\ 
$ -20.75 < M_B-5{\rm log}h < -20.50 $ & $ 9.87 \pm 0.76 \times 10^{-4} $ & $ 1.24 \pm 0.09 \times 10^{-3} $ & $ 1.24 \pm 0.08 \times 10^{-3} $ & $ 1.61 \pm 0.07 \times 10^{-3} $ \\ 
$ -21.00 < M_B-5{\rm log}h < -20.75 $ & $ 7.35 \pm 0.96 \times 10^{-4} $ & $ 8.16 \pm 0.67 \times 10^{-4} $ & $ 9.95 \pm 0.65 \times 10^{-4} $ & $ 1.33 \pm 0.07 \times 10^{-3} $ \\ 
$ -21.25 < M_B-5{\rm log}h < -21.00 $ & $ 3.37 \pm 0.47 \times 10^{-4} $ & $ 5.12 \pm 0.46 \times 10^{-4} $ & $ 6.44 \pm 0.34 \times 10^{-4} $ & $ 8.97 \pm 0.46 \times 10^{-4} $ \\ 
$ -21.50 < M_B-5{\rm log}h < -21.25 $ & $ 1.96 \pm 0.22 \times 10^{-4} $ & $ 2.62 \pm 0.37 \times 10^{-4} $ & $ 4.66 \pm 0.41 \times 10^{-4} $ & $ 5.94 \pm 0.25 \times 10^{-4} $ \\ 
$ -21.75 < M_B-5{\rm log}h < -21.50 $ & $ 3.76 \pm 2.03 \times 10^{-5} $ & $ 1.00 \pm 0.19 \times 10^{-5} $ & $ 2.20 \pm 0.22 \times 10^{-4} $ & $ 3.06 \pm 0.22 \times 10^{-4} $ \\ 
$ -22.00 < M_B-5{\rm log}h < -21.75 $ & $ 2.38 \pm 2.41 \times 10^{-5} $ & $ 4.76 \pm 1.09 \times 10^{-5} $ & $ 9.47 \pm 1.97 \times 10^{-5} $ & $ 1.34 \pm 0.14 \times 10^{-4} $ \\ 
$ -22.25 < M_B-5{\rm log}h < -22.00 $ & -                                & $ 1.19 \pm 0.45 \times 10^{-5} $ & $ 3.21 \pm 0.63 \times 10^{-5} $ & $ 5.30 \pm 0.74 \times 10^{-5} $ \\ 
$ -22.50 < M_B-5{\rm log}h < -22.25 $ & -                                & $ < 1.78 \times 10^{-5}        $ & $ 9.16 \pm 3.08 \times 10^{-6} $ & $ 1.96 \pm 0.54 \times 10^{-5} $ \\ 
$ -22.75 < M_B-5{\rm log}h < -22.50 $ & -                                & $ < 2.20 \times 10^{-5}        $ & $ 1.53 \pm 1.35 \times 10^{-6} $ & $ 4.61 \pm 2.42 \times 10^{-6} $ \\ 
$ -23.00 < M_B-5{\rm log}h < -22.75 $ & -                                & $ < 3.02 \times 10^{-5}        $ & $ 1.53 \pm 1.61 \times 10^{-6} $ & $ 1.15 \pm 1.03 \times 10^{-6} $ \\ 
\enddata
\end{deluxetable}
\begin{deluxetable}{ccccccc}
\tablecolumns{7}
\tabletypesize{\scriptsize}
\tablecaption{Recent  $B$-band luminosity functions of red galaxies\label{table:mlf}}
\tablehead{
\colhead{Survey\tablenotemark{a}}                                        &
\colhead{$z$ range}                                                      &
\colhead{$N_{galaxy}$}                                                   &
\colhead{$M^*_B - 5 {\rm log} h $\tablenotemark{b}}                      &
\colhead{$M_B(10^{-3.5})- 5 {\rm log} h$\tablenotemark{c}}               &
\colhead{$\phi^* (h^3 {\rm Mpc^{-3}~mag^{-1}})       $}                  &
\colhead{$\alpha$\tablenotemark{d}}
}
\startdata
\bootes & $0.20<z<0.40$ &  5983  & $ -19.54 \pm 0.05 $    & $ -21.10 \pm 0.03 $    & $ 8.45 \pm 1.07 \times 10^{-3} $   & $ -0.28 \pm  0.04 $   \\ 
\bootes & $0.40<z<0.60$ & 10983  & $ -19.72 \pm 0.04 $    & $ -21.25 \pm 0.03 $    & $ 7.61 \pm 0.68 \times 10^{-3} $   & $ -0.28 \pm  0.05 $   \\ 
\bootes & $0.60<z<0.80$ & 10817  & $ -20.16 \pm 0.05 $    & $ -21.47 \pm 0.02 $    & $ 5.71 \pm 0.35 \times 10^{-3} $   & $ -0.55 \pm  0.06 $   \\ 
\bootes & $0.80<z<1.00$ & 11816  & $ -20.21 \pm 0.03 $    & $ -21.61 \pm 0.02 $    & $ 6.35 \pm 0.47 \times 10^{-3} $   & $ -0.43 \pm  0.05 $   \\ 
\\
\bootes & $0.20<z<0.40$ &  5983  & $ -19.78 \pm 0.02 $    & $ -21.19 \pm 0.03 $    & $ 7.16 \pm 0.90 \times 10^{-3} $   & $ -0.50 $   \\ 
\bootes & $0.40<z<0.60$ & 10983  & $ -19.92 \pm 0.02 $    & $ -21.30 \pm 0.03 $    & $ 6.60 \pm 0.59 \times 10^{-3} $   & $ -0.50 $   \\ 
\bootes & $0.60<z<0.80$ & 10817  & $ -20.12 \pm 0.02 $    & $ -21.46 \pm 0.02 $    & $ 5.87 \pm 0.36 \times 10^{-3} $   & $ -0.50 $   \\ 
\bootes & $0.80<z<1.00$ & 11816  & $ -20.26 \pm 0.01 $    & $ -21.62 \pm 0.02 $    & $ 6.17 \pm 0.45 \times 10^{-3} $   & $ -0.50 $   \\ 
\\
2dFGRS & $z<0.15$    & 27540  & $ -19.43 \pm  0.05 $    & $ -20.93 $               & $ 9.9 \pm 0.5   \times 10^{-3} $  &  $ -0.54 \pm 0.02 $ \\
\\
COMBO-17  & $0.2<z<0.4$ &   1096  & $ -19.86 \pm  0.16 $  & $ -21.23 $               & $ 6.38 \pm 2.47 \times 10^{-3} $   &  $ -0.50 $   \\
COMBO-17  & $0.4<z<0.6$ &   1179  & $ -20.00 \pm  0.11 $  & $ -21.33 $               & $ 5.82 \pm 0.94 \times 10^{-3} $   &  $ -0.50 $   \\
COMBO-17  & $0.6<z<0.8$ &   1431  & $ -20.33 \pm  0.12 $  & $ -21.62 $               & $ 5.17 \pm 0.20 \times 10^{-3} $   &  $ -0.50 $   \\
COMBO-17  & $0.8<z<1.0$ &   892   & $ -20.41 \pm  0.14 $  & $ -21.53 $               & $ 3.46 \pm 0.15 \times 10^{-3} $   &  $ -0.50 $   \\
COMBO-17  & $1.0<z<1.2$ &   256   & $ -20.81 \pm  0.16 $  & $ -21.44 $               & $ 1.55 \pm 0.34 \times 10^{-3} $   &  $ -0.50 $   \\
\\
DEEP2  & $0.2<z<0.4$ &   109  & $ -20.25 \pm  0.18 $    & $ -21.53 $               & $ 4.97 \pm 0.48 \times 10^{-3} $   &  $ -0.50 $   \\
DEEP2  & $0.4<z<0.6$ &   173  & $ -20.20 \pm  0.12 $    & $ -21.40 $               & $ 4.13 \pm 0.19 \times 10^{-3} $   &  $ -0.50 $   \\
DEEP2  & $0.6<z<0.8$ &   196  & $ -20.42 \pm  0.06 $    & $ -21.60 $               & $ 3.98 \pm 0.31 \times 10^{-3} $   &  $ -0.50 $   \\
DEEP2  & $0.8<z<1.0$ &   535  & $ -20.34 \pm  0.05 $    & $ -21.41 $               & $ 3.13 \pm 0.11 \times 10^{-3} $   &  $ -0.50 $   \\
DEEP2  & $1.0<z<1.2$ &   178  & $ -20.67 \pm  0.08 $    & $ -21.29 $               & $ 1.58 \pm 0.25 \times 10^{-3} $   &  $ -0.50 $   \\
\\
VVDS   & $0.2<z<0.4$ &    65  & $ -20.27 \pm^{0.27}_{0.31} $  & $ -21.66 $         & $ 5.15 \pm 0.64 \times 10^{-3} $   &  $ -0.29 $   \\
VVDS   & $0.4<z<0.6$ &   106  & $ -20.49 \pm^{0.17}_{0.18} $  & $ -21.67 $         & $ 3.12 \pm 0.30 \times 10^{-3} $   &  $ -0.29 $   \\
VVDS   & $0.6<z<0.8$ &   197  & $ -20.22 \pm^{0.09}_{0.10} $  & $ -21.46 $         & $ 3.53 \pm 0.35 \times 10^{-3} $   &  $ -0.29 $   \\
VVDS   & $0.8<z<1.0$ &   164  & $ -20.73 \pm^{0.11}_{0.12} $  & $ -21.76 $         & $ 2.36 \pm 0.18 \times 10^{-3} $   &  $ -0.29 $   \\
VVDS   & $1.0<z<1.2$ &   114  & $ -20.53 \pm^{0.11}_{0.12} $  & $ -21.57 $         & $ 2.39 \pm 0.22 \times 10^{-3} $   &  $ -0.29 $   \\
\\
\enddata
\tablenotetext{a}{\bootes~ (this work), 
2dFGRS \markcite{mad02}(2dF Galaxy Redshift Survey; {Madgwick} {et~al.} 2002), 
COMBO-17 \markcite{bel04,fab05}(Classifying Objects by Medium-Band Observations - a spectrophotometric 17-filter survey; {Bell} {et~al.} 2004; {Faber} {et~al.} 2005), 
DEEP2 \markcite{wil05,fab05}(Deep Extragalactic Evolution Probe 2; {Willmer} {et~al.} 2005; {Faber} {et~al.} 2005), 
VVDS \markcite{zuc05}(VIRMOS-VLT Deep Survey; {Zucca} {et~al.} 2005)}
\tablenotetext{b}{We have adopted $B_J-B=0.15$ and $B_{\rm Vega}=B_{\rm AB}$. Uncertainties are as published and may not account
for the contribution of large-scale structure.}
\tablenotetext{c}{We have not determined uncertainties for other surveys, but uncertainties for $M(10^{-3.5} h^3~{\rm Mpc^{-3}~mag^{-1}})$
are probably comparable to those for $M^*$}.
\tablenotetext{d}{Values of $\alpha$ without uncertainties denote Schechter function fits where $\alpha$ was fixed.}
\end{deluxetable}

\begin{deluxetable}{ccccccc}
\tablecolumns{7}
\tabletypesize{\scriptsize}
\tablecaption{$B$-band Luminosity Density of Red Galaxies\label{table:jbtab}}
\tablehead{
\colhead{$z$ range}                                                        &
\multicolumn{5}{c}{Luminosity Density $j_B (10^{7} h~L_{\odot}~{\rm Mpc^{-3}})$\tablenotemark{ab}}    &
\colhead{$\alpha$}  \\
\colhead{} &
\colhead{All Luminosities} &
\colhead{$M_B<M^*+1.0$} &
\colhead{$M_B<M^*$} &
\colhead{$M_B<M^*-1.0$} &
\colhead{$M_B<M^*-1.5$\tablenotemark{c}} &
\colhead{} 
}
\startdata
$0.20<z<0.40$  & $  8.08 \pm  1.02 $   & $  7.05 \pm  0.89 $   & $  5.09 \pm  0.64 $   & $  1.70 \pm  0.21 $   & $  0.50 \pm  0.06 $   & $ -0.28 \pm  0.04 $   \\ 
$0.40<z<0.60$  & $  8.47 \pm  0.75 $   & $  7.49 \pm  0.67 $   & $  5.42 \pm  0.48 $   & $  1.80 \pm  0.16 $   & $  0.53 \pm  0.05 $   & $ -0.28 \pm  0.05 $   \\ 
$0.60<z<0.80$  & $  9.18 \pm  0.56 $   & $  7.62 \pm  0.47 $   & $  5.03 \pm  0.31 $   & $  1.44 \pm  0.09 $   & $  0.39 \pm  0.02 $   & $ -0.55 \pm  0.06 $   \\ 
$0.80<z<1.00$  & $  10.7 \pm   0.8 $   & $  9.25 \pm  0.68 $   & $  6.37 \pm  0.47 $   & $  1.95 \pm  0.14 $   & $  0.55 \pm  0.04 $   & $ -0.43 \pm  0.05 $   \\ 
\\
$0.20<z<0.40$  & $  8.29 \pm  1.05 $   & $  6.85 \pm  0.86 $   & $  4.60 \pm  0.58 $   & $  1.36 \pm  0.17 $   & $  0.37 \pm  0.05 $   & $ -0.50 $   \\ 
$0.40<z<0.60$  & $  8.57 \pm  0.76 $   & $  7.18 \pm  0.64 $   & $  4.82 \pm  0.43 $   & $  1.42 \pm  0.13 $   & $  0.39 \pm  0.03 $   & $ -0.50 $   \\ 
$0.60<z<0.80$  & $  9.10 \pm  0.56 $   & $  7.68 \pm  0.47 $   & $  5.16 \pm  0.32 $   & $  1.52 \pm  0.09 $   & $  0.42 \pm  0.03 $   & $ -0.50 $   \\ 
$0.80<z<1.00$  & $  10.8 \pm   0.8 $   & $  9.19 \pm  0.68 $   & $  6.17 \pm  0.45 $   & $  1.82 \pm  0.13 $   & $  0.50 \pm  0.04 $   & $ -0.50 $   \\ 
\enddata
\tablenotetext{a}{The Sun's absolute magnitude is $M_B=5.48$ \markcite{bes98}({Bessell}, {Castelli}, \& {Plez} 1998).}
\tablenotetext{b}{When integrating over portions of the Schechter function, the resulting $j_B$ values
can be very sensitive to the assumed or measured values of $M^*$ and $\alpha$. For example, when 
$\alpha$ is fixed  $j_B(M_B<M^*-1)/j_B(M_B<M^*+1)$ is a constant irrespective of the redshift or $\phi^*$.}
\tablenotetext{c}{As discussed in \S\ref{sec:msec}, small errors in assumed or measured values of 
$M^*$ and $\alpha$ can produce large errors in the measured luminosity density of $M_B<M^*-1.5$ red galaxies.}
\end{deluxetable}

\end{document}